\documentclass[twocolumn]{aastex631}
\usepackage{graphicx} % Required for inserting images
\usepackage{hyperref}
\usepackage{siunitx}
\usepackage{longtable}
\usepackage{soul}
\usepackage{amsmath}
\newcommand{\figr}[1]{Figure~\ref{fig:#1}}
\newcommand{\secr}[1]{Section~\ref{sec:#1}}
\newcommand{\eqr}[1]{Eq.~\ref{eq:#1}}
\newcommand{\tabr}[1]{\mbox{Table~\ref{tab:#1}}}

\newcommand{\ik}{{\it Kepler}}
\newcommand{\kt}{{\it K2}}
\newcommand{\tess}{{\it TESS}}

\begin{document}
\setlength{\abovedisplayskip}{5px}
\setlength{\belowdisplayskip}{5px}
%%%%%%%%%%%%%%%%%%%%%%%%%%%%%%%%%%%%%%%%%%%%%%%%%%%%%%%%%%%%%%%%%%%%%%%%%%%%%
\shorttitle{Temporal Atmospheric Variability of Hot Jupiters}
\shortauthors{Li \& Shporer}
%%%%%%%%%%%%%%%%%%%%%%%%%%%%%%%%%%%%%%%%%%%%%%%%%%%%%%%%%%%%%%%%%%%%%%%%%%%%%

%%%%%%%%%%%%%%%%%%%%%%%%%%%%%%%%%%%%%%%%%%%%%%%%%%%%%%%%%%%%%%%%%%%%%%%%%%%%%
\title{A Search for Temporal Atmospheric Variability of \textit{Kepler} Hot Jupiters}

\author{Canis Li}
\affil{\textit{Valley Christian High School, San Jose, CA 95111, USA}}

\author{Avi Shporer}
\affil{\textit{Department of Physics and Kavli Institute for Astrophysics and Space Research, Massachusetts Institute of Technology, Cambridge, MA 02139, USA}}

%\date{November 8, 2023}
%%%%%%%%%%%%%%%%%%%%%%%%%%%%%%%%%%%%%%%%%%%%%%%%%%%%%%%%%%%%%%%%%%%%%%%%%%%%%

% \maketitle
%%%%%%%%%%%%%%%%%%%%%%%%%%%%%%%%%%%%%%%%%%%%%%%%%%%%%%%%%%%%%%%%%%%%%%%%%%%%%
\begin{abstract}
    We perform a systematic search for atmospheric variability in short-period gas-giant planets (hot Jupiters) observed by the \ik\ mission, by looking for temporal variability of their secondary eclipse depths.
    This is motivated by a recent detection of a decrease in the dayside brightness of KELT-1~b between TESS Sectors 17 and 57, separated by about 3 years.
    We fit the \ik\ light curves of 53 hot Jupiters and measure their secondary eclipse depths during individual \textit{Kepler} quarters and 4-quarter windows. We detect the secondary eclipses in individual quarters or four-quarter windows for 17 out of the 53 systems.
    In those 17 systems we do not detect statistically significant astrophysical variation in the secondary eclipse depths.  
    We show that the data is sensitive to the variability seen for KELT-1~b in TESS data.
    Therefore, the absence of detected secondary eclipse variability in \textit{Kepler} data suggests that the atmospheric variability in KELT-1~b is not common.
    In addition, several of the 53 targets we investigated display variability in their transit depths with a period of 4 quarters (1 year). This instrumental signal is likely present in the light curves of other transiting planets we did not analyze and other variable stars observed by \ik.   
    Finally, we find that Kepler-488~b has a secondary eclipse depth that is unphysically large for a planet, and thus is likely a misclassified red dwarf.  
\end{abstract}
%%%%%%%%%%%%%%%%%%%%%%%%%%%%%%%%%%%%%%%%%%%%%%%%%%%%%%%%%%%%%%%%%%%%%%%%%%%%%


%%%%%%%%%%%%%%%%%%%%%%%%%%%%%%%%%%%%%%%%%%%%%%%%%%%%%%%%%%%%%%%%%%%%%%%%%%%%%
\section{Introduction}
%%%%%%%%%%%%%%%%%%%%%%%%%%%%%%%%%%%%%%%%%%%%%%%%%%%%%%%%%%%%%%%%%%%%%%%%%%%%%

The availability of high-quality continuous time-series photometry in visible light for a large number of stars, provided by recent space missions (e.g., {\it CoRoT}, \ik, \kt, \tess) has revolutionized the study of transiting exoplanets in particular, and more broadly the study of stellar photometric variability \cite[e.g.,][]{chaplin14, mcquillan14, kirk16, shporer17, prsa22}. While designed to detect small transiting planets, data provided by these missions are sensitive also to the shallow secondary eclipse --- when the planet orbits behind its host star and is occulted by it --- of short-period gas-giant planets, so-called hot Jupiters \cite[e.g.,][]{borucki09, morris13, parviainen13, angerhausen15}. The visible light secondary eclipse depth can reach an order of 100 parts-per-million (ppm), similar to the transit depth of Earth-like planets transiting Sun-like stars. Visible light originating from a hot Jupiter can be reflected light or thermal emission, or a combination of both. Therefore, a measurement of the visible light secondary eclipse depth provides constraints on the planet's atmospheric albedo and temperature \cite[e.g.,][]{demory11, shporer14, angerhausen15, wong20, wong21}, which in turn gives insight into the atmosphere's structure and content \citep[e.g.,][]{demory13, parmentier16, showman20}.

The continuous four-year time span of the \ik\ mission allows the search for variability in hot Jupiter atmospheric properties over that time scale.  
\citet{armstrong16} and \citet{jackson19} used \ik\ data to study atmospheric variability of HAT-P-7~b/Kepler-2~b and Kepler-76~b, respectively, by monitoring the shape of the full orbital phase curve, and claimed detection of variability in the cloud coverage.
However, the detection of variability for HAT-P-7~b was shown by \citet{lally22} to be likely due to stellar super-granulation and not planetary atmospheric variability.

We report here the results of a search for variability in atmospheres of hot Jupiters observed by \ik\ by looking for variability in the secondary eclipse depths between \ik\ quarters. Our study is motivated by the $286 \pm 65$ ppm ($ 4.4 \sigma$) decrease in the secondary eclipse depth of the brown dwarf KELT-1~b \citep{siverd12} between two \tess\ data sets, Sectors 17 and 57, obtained 3 years apart, recently reported by \cite{parviainen23}. This variability is thought to be due to changes in the silicate cloud coverage on the dayside of KELT-1~b.

We describe our data analysis in \secr{dataanalysis}, and the results in \secr{results}. In \secr{discussion} we provide a discussion of our results, and we present our conclusions in \secr{summary}.

%%%%%%%%%%%%%%%%%%%%%%%%%%%%%%%%%%%%%%%%%%%%%%%%%%%%%%%%%%%%%%%%%%%%%%%%%%%%%
\section{Analysis of \textit{Kepler} light curves}
\label{sec:dataanalysis}
%%%%%%%%%%%%%%%%%%%%%%%%%%%%%%%%%%%%%%%%%%%%%%%%%%%%%%%%%%%%%%%%%%%%%%%%%%%%%

In this section, we describe our methodology for analyzing the \textit{Kepler} data, including target selection, transit and secondary eclipse model fitting, and looking for temporal variability.

%----------------------------------------------------------------------------
\subsection{Target selection} 
\label{sec:targetselection}
%----------------------------------------------------------------------------

As noted above, the depth of a planetary secondary eclipse in visible light is up to the order of 100 ppm. Therefore we first compile a list of transiting hot Jupiters observed by \ik\ where such a shallow eclipse might be detectable in individual \ik\ quarters, each with a duration of about 90 days. We select star-planet systems with an orbital period shorter than 15 days ($P<15$ days), planet radius larger than 5 Earth radii ($R_p > 5~R_{\earth}$), and host stars brighter than 15.0 mag in the \ik\ band ($Kp<$ 15.0 mag). We apply these criteria to the \ik\ systems listed on the NASA Exoplanet Archive\footnote{\url{https://exoplanetarchive.ipac.caltech.edu}\label{nea_table}}.
\tabr{kepler_targets} lists the 53 planets satisfying our selection criteria, sorted by \textit{Kepler} planet number. 

To estimate an upper bound for the secondary eclipse depths, ${F_p}/{F_*}$, of these 53 planets, we use the preexisting default parameter values on the NASA Exoplanet Archive to compute two values. First, we calculate the depth assuming it is only due to reflected light from the planet:
\begin{equation}{F_p}/{F_*} = A_g \left(\frac{R_p}{a}\right)^2,\label{eq:albedo}\end{equation}
where $R_p$ and $a$ are the planetary radius and orbital semi-major axis, respectively, and $A_g$ is the geometric albedo in visible light. To estimate the maximum secondary eclipse depth we assume $A_g=1.0$. While a geometric albedo can in principle be larger than 1.0 (caused for example by non-isotropic scattering), measured geometric albedos for hot Jupiters are well below 1.0 \citep[e.g.,][]{angerhausen15, esteves15, wong20, wong21}. Next, we compute the depth assuming it is entirely due to thermal emission from the planet, treating it and the star as blackbodies. Assuming no heat circulation between the planet's dayside and nightside hemispheres, the planetary dayside equilibrium temperature is
\begin{equation}T_{\text{eq}} = \frac{T_*}{\sqrt[4]{2}} \sqrt{\frac{R_*}{a}},\label{eq:T_eq_dayside}\end{equation}
where $T_*$ and $R_*$ are the stellar effective temperature and radius, respectively.
We obtain the depth by dividing the planet's thermal radiation integrated under the \textit{Kepler} passband by that of the star. 

For the 53 systems, the predicted maximum depth due to thermal emission in the \ik\ passband is an order of magnitude smaller than the depth due to reflected light, with the exception of Kepler-13~b where thermal emission is about 5 times smaller. In \tabr{kepler_targets}, the rightmost column denotes the predicted depth due to reflected light. These values serve as a sanity check for making sure our fitted secondary eclipse depths are physical.

%----------------------------------------------------------------------------
\subsection{Observations}
\label{sec:obs}
%----------------------------------------------------------------------------

For each of the 53 selected targets, we use \textit{Kepler} quarters 1 through 17 of the Presearch Data Conditioning Simple Aperture Photometry (PDC-SAP) long-cadence photometry \citep{smith12, stumpe12}, with an effective integration time of 30 minutes\footnote{Available online through MAST: \dataset[10.17909/T9488N]{http://dx.doi.org/10.17909/T9488N}}. To test our results we have also analyzed the detrended light curves from the Data Validation time series version of the data, obtained by applying a median filter to the PDC-SAP light curve \citep{twicken18}, which showed consistent results with that of the PDC-SAP data.

%----------------------------------------------------------------------------
\subsection{Light curve fitting} 
\label{sec:light_curve_fitting}
%----------------------------------------------------------------------------

While our goal is to look for variability in the secondary eclipse depth, we simultaneously fit the secondary eclipse and the transit. We do that in order to use the transit depth as a control for identifying cases where depth variations arise from instrumental processes affecting the amplitude of all variability signals in the light curve, such as varying quality of correction for blending between quarters. In such cases, both the secondary eclipse and transit depths will vary in the same manner. 

We fit each quarter's light curve using Bayesian inference and Monte Carlo Markov Chain (MCMC) analysis. We employ the \textsc{SecondaryEclipseLightCurve} model from the \textsc{exoplanet} Python package \citep{exoplanet}, which models dips at the planet's transit and secondary eclipse epochs computed by \textsc{starry} \citep{starry}. We model light curves of multi-planetary systems by modeling each planet's light curve independently, ignoring the influence of other planets, and summing them together.    

We include a Gaussian process (GP) in our fitted model, to address photometric variability beyond the transit and secondary eclipse. We perform GP regression (see, e.g., \citealt{aigrain23}) using the stochastically-driven, damped harmonic oscillator kernel implemented by \textsc{celerite2} \citep{celerite2}. It can robustly account for jumps in flux, temporal gaps, as well as both non-periodic and quasi-periodic variations resulting for example from stellar activity combined with stellar rotation. We characterize the kernel with two free parameters, the undamped oscillation period, $\rho_\text{GP}$, and the oscillations standard deviation, $\sigma_\text{GP}$. We set the quality factor $Q$ to be $\frac{1}{\sqrt{2}}$. The power spectral density of this kernel is shown below: 
\begin{equation}
\begin{split}
    S(\omega) = \sqrt{\frac{2}{\pi}} \frac{S_0\,\omega_0^4}
{(\omega^2-{\omega_0}^2)^2 + {\omega_0}^2\,\omega^2/Q^2}, \\\text{where}\quad\rho_\text{GP}=\frac{2\pi}{\omega_0}, \,\sigma_\text{GP}=\sqrt{S_0\omega_0 Q}.
\end{split}
\end{equation}
To account for excess variance, we also fit for a jitter term $s$, which effectively increases the individual flux measurement's variance, $\sigma_n^2$, to $\sigma_n^2+s^2$.

The free parameters we fit for are the orbital period, $P$, the transit epoch, $t_0$, the planet-to-star radius ratio, $R_p/R_*$, the scaled semi-major axis, $a/R_*$, the impact parameter, $b$, the stellar quadratic limb-darkening parameters, $q_1, q_2$, and secondary eclipse depth,  ${F_p}/{F_*}$. We employ the \citet{kipping13b} limb-darkening parameterization, which ensures the star's intensity profile decreases monotonically from the center and remains positive. The transit depth is derived analytically using the \citet{mandelagol} formalism.

For most of our targets, because they are hot Jupiters their orbits can be assumed to be circular, meaning the secondary eclipse occurs exactly halfway between transit times. Even if the orbital eccentricity is non-zero, it is expected that in most cases it is too small to be measured using the shape of the secondary eclipse (specifically its phase and duration relative to those of the transit), especially with only one or four \ik\ quarters. For the few systems that have previously been reported to have a non-circular orbit (e.g., Kepler-63~b) and/or seem to have their secondary eclipse phase deviating from halfway between transits (Kepler-488~b), we fit for $\sqrt{e}\sin\omega$ and $\sqrt{e}\cos\omega$, where $e$ is the eccentricity and $\omega$ is the argument of periastron. For the rest of the systems, we fix $e=0$. 

We can neglect the light travel time delay between transit and secondary eclipse events due to the proximity of the orbits of our target planets to their host stars, making the delay smaller than the typical uncertainty on the mid secondary eclipse time. 

The prior distributions we employ for each parameter are listed in \tabr{priors}. Except for the period, transit epoch, and GP amplitude, the priors are non-informative. We allow the secondary eclipse depth to become negative, although that is unphysical, in order to allow for non-detections of the eclipse. 

When analyzing each quarter's light curve, we first optimize the light curve and GP model for the maximum a posteriori (MAP) solution for the free parameters. These values correspond to a mode in the posterior distribution. The starting point for our optimizer is the star-planet system's default parameter set provided in the NASA Exoplanet Archive Planetary Systems table\footnote{We substituted parameter values from the \ik\ Objects of Interest DR24 table \citep{seader15} whenever they were unavailable in the default parameter set of the NASA Exoplanet Archive table.}\citep{NEA_planetarysystems}. We clip $5\sigma$ residuals and then perform the same process to obtain a new MAP solution. We then use this MAP solution as the MCMC starting point.

For systems with active host stars, the amplitude of the nuisance signal can be comparable to or even larger than the transit depth. In these cases, reliably fitting the eclipses (transits and secondary eclipses) and the GP simultaneously is difficult since the GP might overfit, especially during individual quarters where there are few transit events. To minimize these occurrences, we perform three successive optimizations to obtain the MAP solution (described above) for all of our systems: First, we optimize the GP hyperparameters, leaving the model light curve parameters fixed at their starting values. From there, we do the same but vice versa, leaving the optimized GP hyperparameters constant. Finally, we optimize both simultaneously. Since the starting point taken from the NASA Exoplanet Archive parameters table is already a good estimate for the model light curve, optimizing just the GP first prevents the optimizer from approaching a mode where the transits are overfitted. 

In order to assess the level of secondary eclipse depth variability on year-long timescales, instead of between \ik\ quarters, we also performed fits to sliding four-quarter windows of the full light curve.
This gives 14 data points, each with a higher signal-to-noise ratio compared to that of single quarters. 

%----------------------------------------------------------------------------
\begin{table*}[htb!]
\centering
\begin{tabular}{lcrr}
\hline 
\hline \\ [-2ex]
Parameter & Prior & Description\\
\hline
\multicolumn{1}{l}{Planet and star} \\
$P$ & $\mathcal{N}(P_\text{NEA}, \SI{0.1})$ & Orbital period [d]\\
$t_0$ & $\mathcal{N}(t_{0, \text{NEA}}, \SI{0.1})$ & Transit epoch [bkjd]\\
$R_p/R_*$ & $\mathcal{U}(0, 1)$ & Planet-to-star radius ratio\\
$b$ & $\mathcal{U}(0, 1)$ & Impact parameter\\
$a/R_*$ & $\mathcal{U}(1, \infty)$ & Scaled semi-major axis\\
$e$ & 0 (fixed)\tablenotemark{a} & Eccentricity \\
$q_1, q_2$ & $\mathcal{U}(0,1),\ \mathcal{U}(0,1)$ & Stellar limb-darkening\\
%$q_{1, p}, q_{2, p}$ & $0, 0$ ( fixed) & Planet limb-darkening\\
${F_p}/{F_*}$ & $\mathcal{U}(-1000, 1000)$ & Secondary eclipse depth [ppm] \\
\multicolumn{1}{l}{Detrending} \\
$\rho_\text{GP}$   &  $\mathcal{U}(1, 100)$ & Undamped period of oscillations [d] \\
$\sigma_\text{GP}$ &  $\mathcal{U}(0, 1.5\sigma_\text{LC})$  & Standard deviation of oscillations [ppm] \\
$s$ & $\mathcal{U}(0, 500)$ & Jitter [ppm]\\
$f_0$ & $\mathcal{N}(0, 100)$ & Normalized relative flux offset [ppm]\\
% (log) Error of lightcurve, $\log{\sigma_\text{lc}}$  &  nontrivial* \\
\hline
\end{tabular}
\caption{Priors used for each light curve model parameter described in \secr{light_curve_fitting}. Note that $P_\text{NEA}$ and $t_{0,\text{NEA}}$ refer to the transit ephemeris obtained from the NASA Exoplanet Archive Planetary Systems table \citep{NEA_planetarysystems}, $\sigma_\text{LC}$ represents the standard deviation of the out-of-transit light curve,
$\mathcal{U}(a, b)$ denotes a uniform distribution from $a$ to $b$, and $\mathcal{N}(\mu, \sigma)$ denotes a normal distribution with mean $\mu$ and variance $\sigma^2$.}
\vspace{-5px}
\tablenotetext{a}{With the exception of two systems we deemed appropriate to fit for eccentricity: Kepler-63~b and Kepler-488~b.}
\label{tab:priors}
\end{table*}
%----------------------------------------------------------------------------

%----------------------------------------------------------------------------
\subsection{Searching for secondary eclipse variability}
\label{sec:variability_search}
%----------------------------------------------------------------------------

We look for variability in systems where the secondary eclipse is detected at a $3\sigma$ statistical significance in at least one quarter, or, one 4-quarter window. Since there are at most 17 measurements per object, the likelihood of a $3\sigma$ event due to Gaussian noise is low\footnote{In a Gaussian distribution, 0.135\% (or 1 part in 741) are $3\sigma$ or more above the mean.}. The excluded systems, where the secondary eclipse depth is not detected or only marginally detected in all quarters cannot display statistically significant variability. 

For systems with a detected secondary eclipse in at least one quarter or 4-quarter window, we compute several statistical quantities. Those include the root mean square (RMS) of the quarterly secondary eclipse depths, the RMS divided by the average error of the eclipse depth measurements, and the reduced chi-squared statistic, $\chi^2_\nu$, along with its associated p-value, $P(\chi^2_\nu)$, calculated by expecting a constant eclipse depth equal to the mean depth across all available quarters.

%%%%%%%%%%%%%%%%%%%%%%%%%%%%%%%%%%%%%%%%%%%%%%%%%%%%%%%%%%%%%%%%%%%%%%%%%%%%%
\section{Results}
\label{sec:results}
%%%%%%%%%%%%%%%%%%%%%%%%%%%%%%%%%%%%%%%%%%%%%%%%%%%%%%%%%%%%%%%%%%%%%%%%%%%%%

\tabr{kepler_targets_with_resolved_eclipses} lists the 17 systems for which we identified a secondary eclipse measured at a statistical significance of at least 3$\sigma$ in at least one \ik\ quarter or one 4-quarter window, along with some of the star-planet system properties and statistics of the quarterly secondary eclipse depth time series, described in the previous paragraph. 
We present the results for these 17 planets visually in Figures 1, 2, and 3. For each system, we plot the depths of the secondary eclipses in individual \ik\ quarters in the top panel and the transit depths in the bottom panel. The depths are plotted in relative flux, in ppm, as a function of quarter number. The blue line marks the depths taking 4 contiguous \ik\ quarters combined. The depth derived by combining all available \ik\ data is plotted as a solid horizontal red line, with $\pm 1\sigma$ uncertainty indicated as the shaded area.

For the majority of these 17 objects, no statistically significant variability in the secondary eclipse depth is observed. For a few of them, we see marginal variability in 1--2 quarters where the secondary eclipse depth deviates by at least 3.0$\sigma$ from the mean.  
In the following paragraphs we briefly describe systems showing a potential variability in their secondary eclipse depth or transit depth.

\begin{table*}[htb!]
% \addtolength{\leftskip}{-1.25cm}
\centering
\begin{tabular}{lrrrrrrccc}
\hline 
\hline \\ [-2ex]
Planet & $P\,\,$& $R_p\,\,$ & $Kp\,\,$ & RMS\, & $\text{RMS}/\langle\sigma_\text{err}\rangle$ & $\chi^2_\nu$ & $P(\chi^2_\nu)$ \\
name & [d]\,\, & [R$_\oplus$] & [mag] & [ppm] &  &  & \\
\hline \\ [-2ex]
Kepler-1 b & 2.47 & 15.2 & 11.34 & 5.50 & 0.637 & 0.584 & 0.869 \\
Kepler-2 b & 2.20 & 16.9 & 10.46 & 8.19 & 1.347 & 2.224 & 0.003 \\
Kepler-5 b & 3.55 & 16.0 & 13.37 & 12.83 & 1.105 & 1.347 & 0.158 \\
Kepler-6 b & 3.23 & 14.6 & 13.30 & 12.59 & 0.963 & 1.134 & 0.324 \\
Kepler-7 b & 4.89 & 18.2 & 12.88 & 13.79 & 1.125 & 1.507 & 0.087 \\
Kepler-8 b & 3.52 & 15.9 & 13.56 & 14.57 & 0.715 & 0.572 & 0.907 \\
Kepler-12 b & 4.44 & 19.7 & 13.44 & 18.15 & 1.026 & 1.315 & 0.195 \\
Kepler-13 b & 1.76 & 16.9 & 9.96 & 2.84 & 0.906 & 0.936 & 0.526 \\
Kepler-15 b & 4.94 & 10.8 & 13.76 & 27.09 & 1.278 & 1.751 & 0.031 \\
Kepler-17 b & 1.49 & 14.7 & 14.14 & 32.51 & 1.162 & 1.436 & 0.134 \\
Kepler-18 c & 7.64 & 5.5 & 13.55 & 32.51 & 1.432 & 2.347 & 0.002 \\
Kepler-41 b & 1.86 & 14.5 & 14.46 & 24.25 & 1.110 & 1.511 & 0.086 \\
Kepler-76 b & 1.54 & 15.2 & 13.31 & 36.48 & 0.834 & 0.756 & 0.708 \\
Kepler-412 b & 1.72 & 15.0 & 14.31 & 21.34 & 0.960 & 0.998 & 0.456 \\
Kepler-427 b & 10.29 & 13.8 & 14.22 & 44.22 & 1.465 & 2.724 & 0.000 \\
Kepler-488 b & 3.12 & 15.8 & 14.80 & 67.18 & 0.624 & 0.929 & 0.530 \\
Kepler-1658 b & 3.85 & 12.0 & 11.43 & 19.93 & 1.203 & 1.579 & 0.083 \\
\hline \\ [-2ex]
\end{tabular}
\caption{Table of \textit{Kepler} systems for which we measured the secondary eclipse in at least one quarter or one 4-quarter window at a significance of $\geq3\sigma$. Columns include, from left to right, the planet name, orbital period, planet radius, host star brightness, root mean square (RMS) of the secondary eclipse depths by quarter, the RMS divided by the average secondary eclipse depth error, and the reduced $\chi^2$ and its associated p-value, assuming a constant depth equal to the mean depth (See \secr{variability_search}).
}
\label{tab:kepler_targets_with_resolved_eclipses}
\end{table*}

%()()()()()()()()()()()()()()()()()()()()()()()()()()()()()()()()()()()()()()()()
\begin{figure*}[htb!]
\gridline{\fig{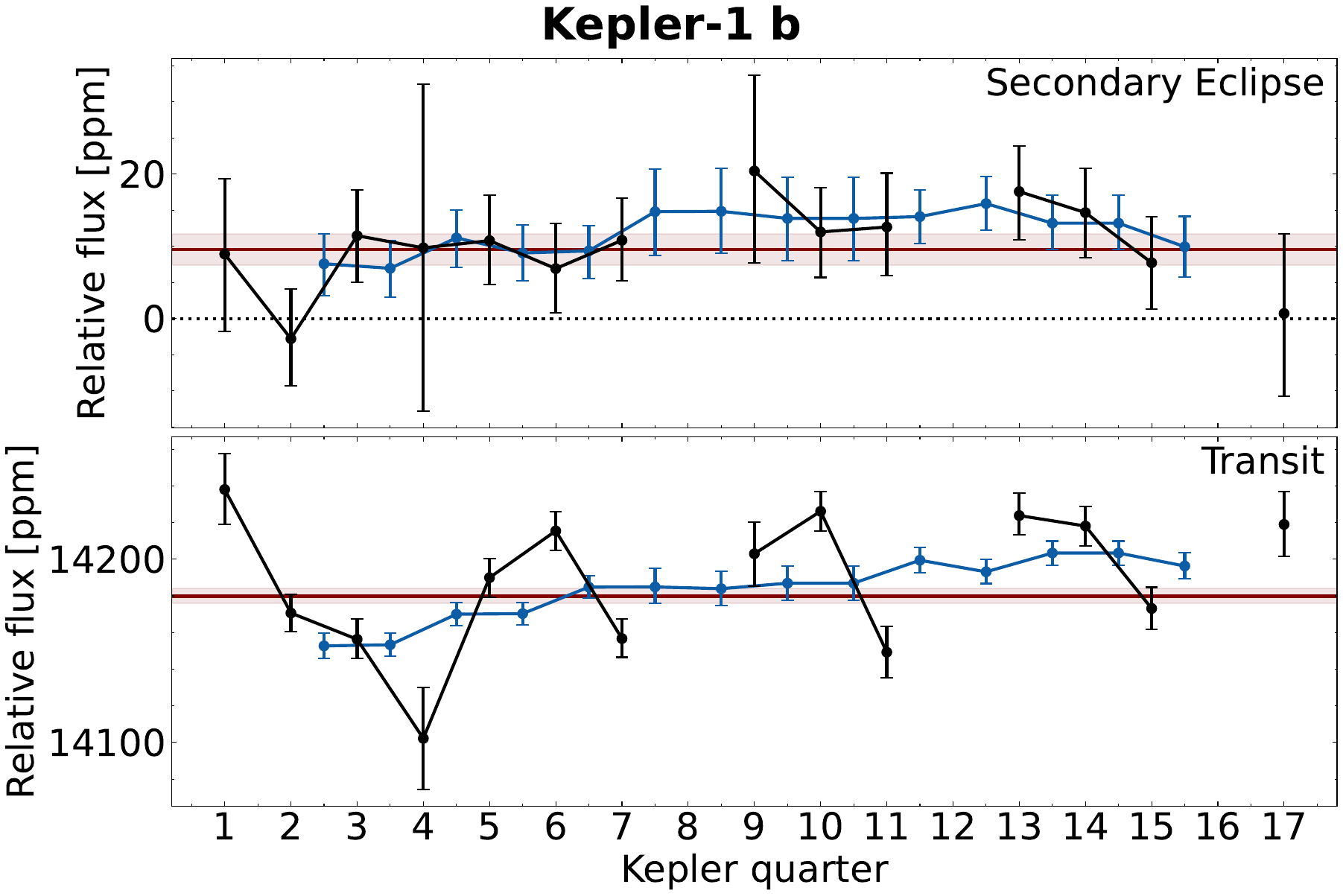}{0.5\textwidth}{}
\fig{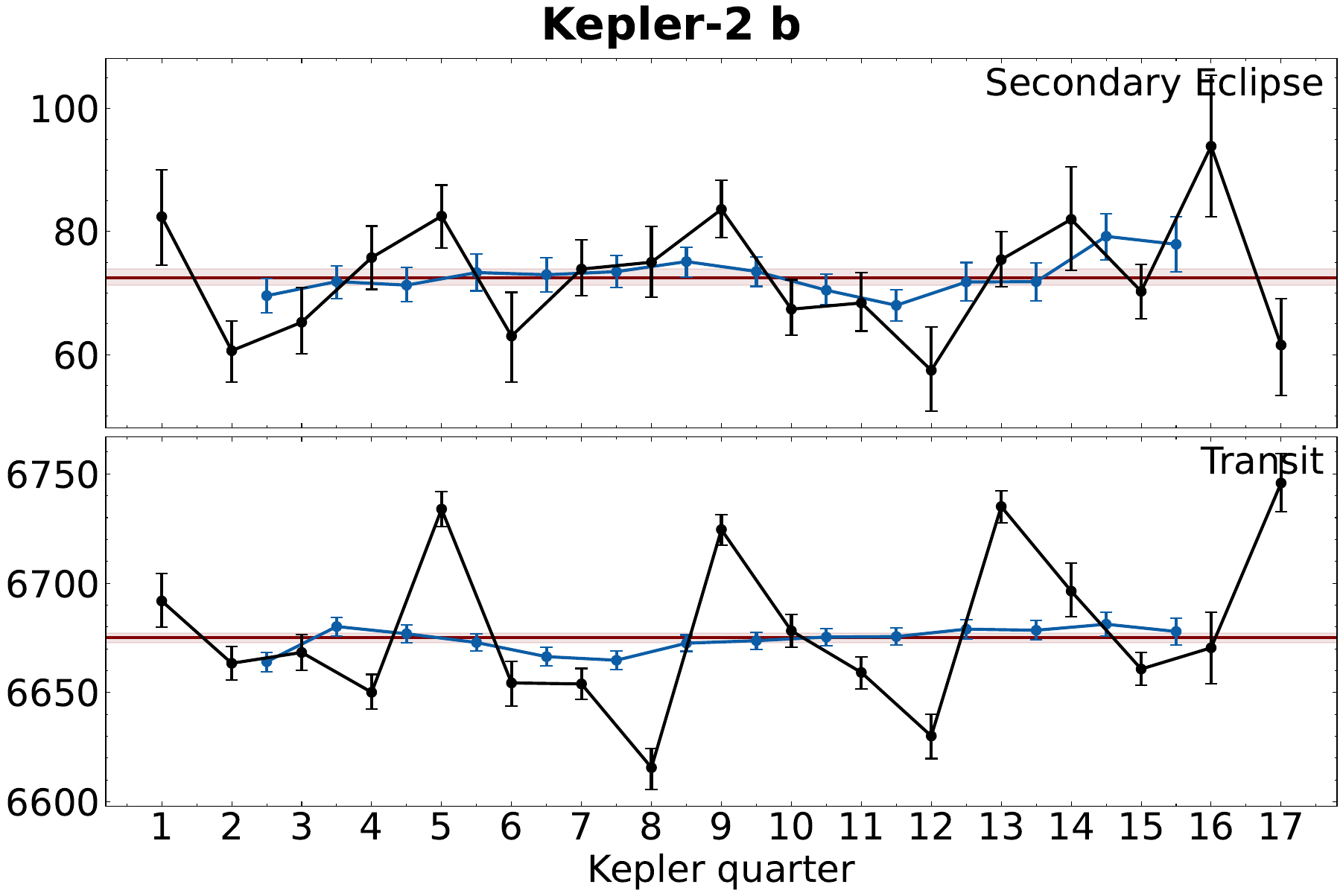}{0.5\textwidth}{}}\vspace{-20px}
\gridline{\fig{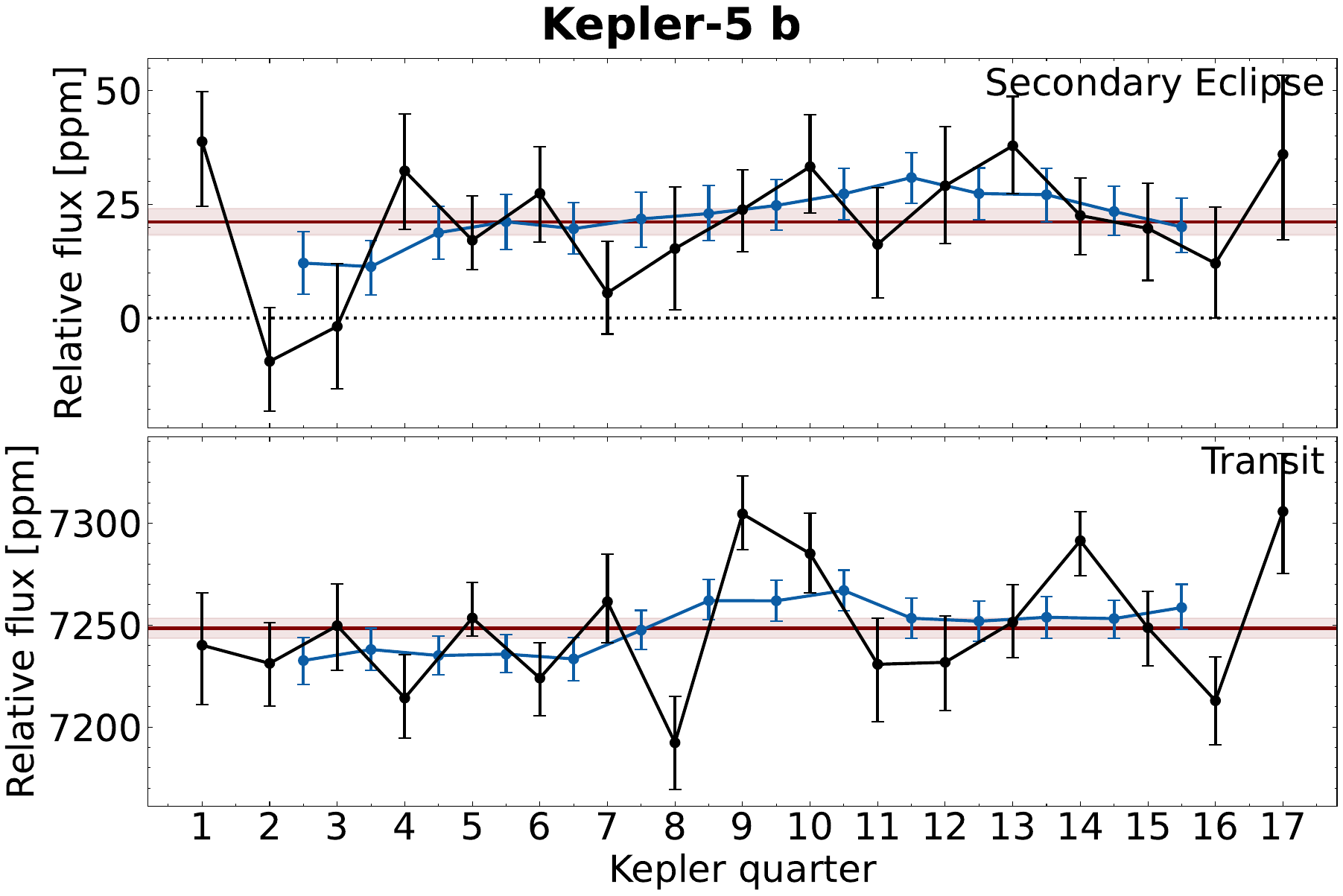}{0.5\textwidth}{}
\fig{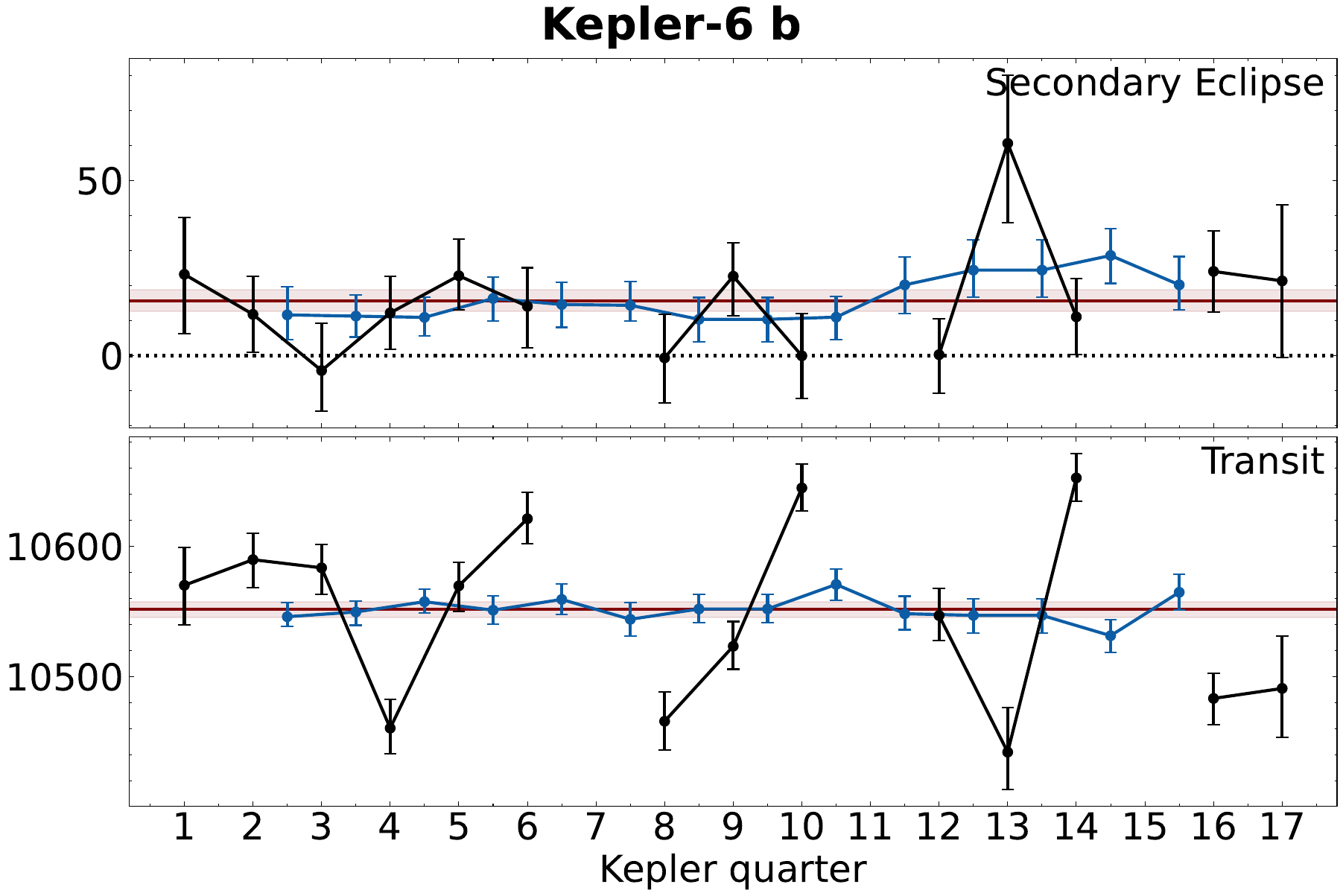}{0.5\textwidth}{}}\vspace{-20px}
\gridline{\fig{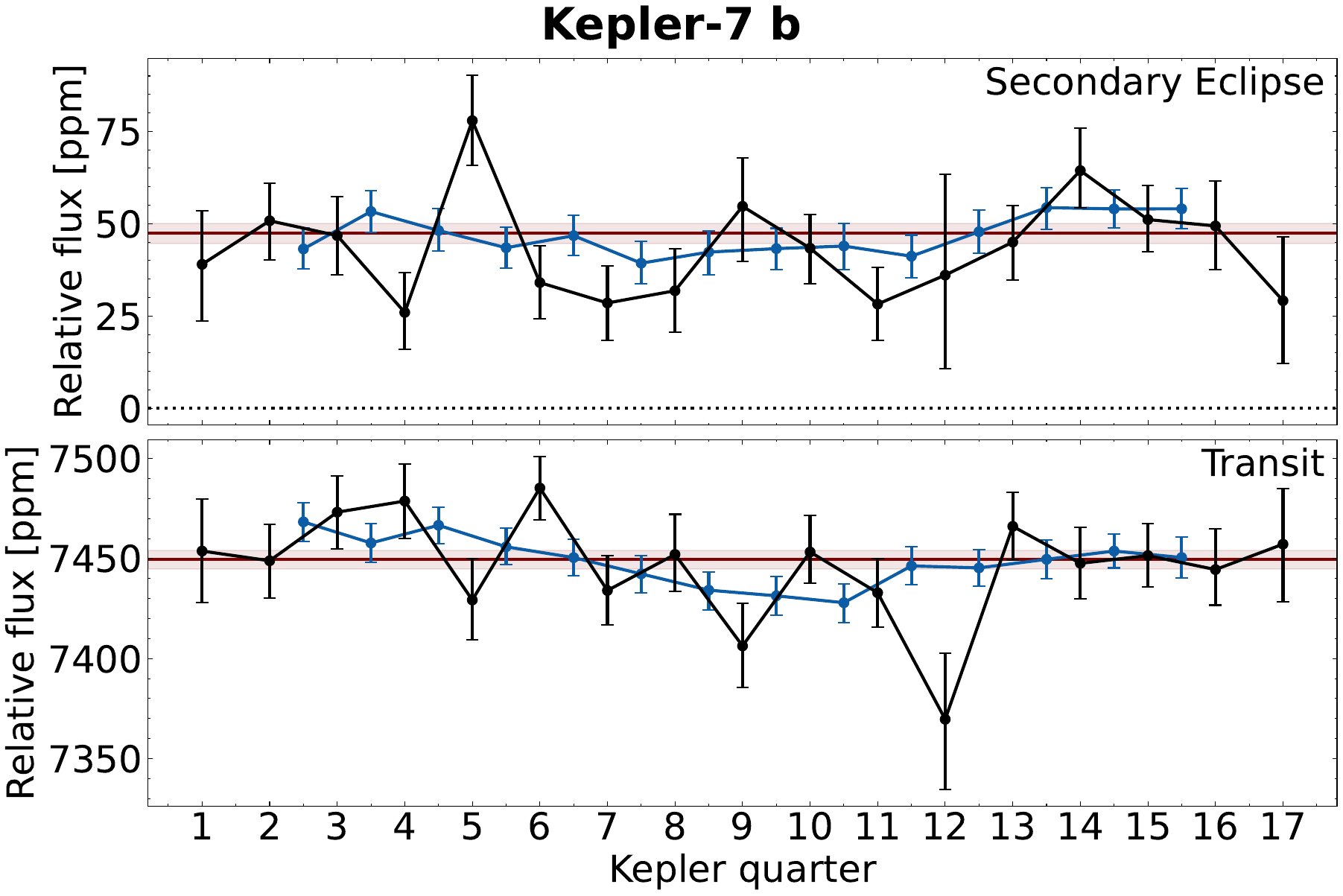}{0.5\textwidth}{}
\fig{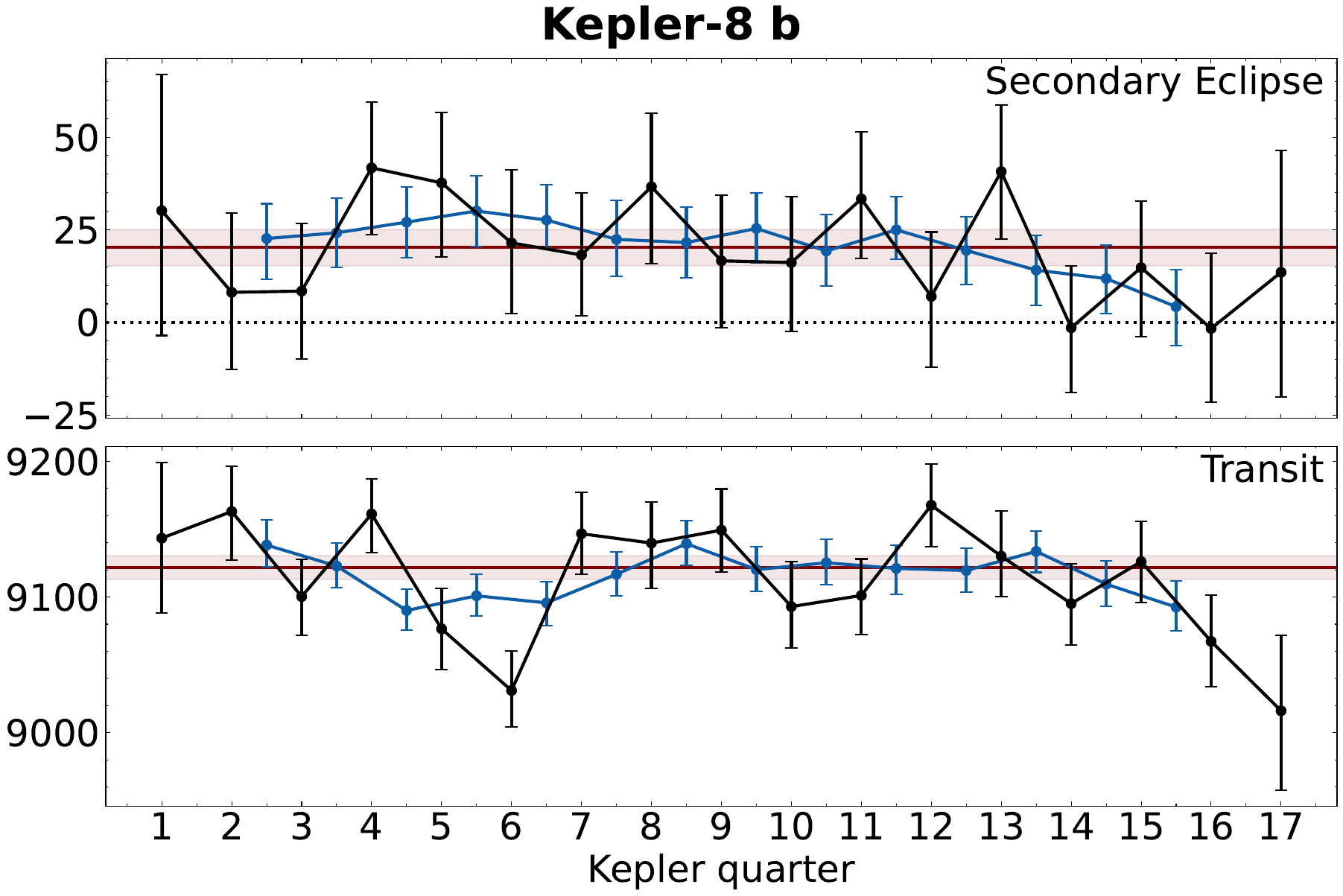}{0.5\textwidth}{}}\vspace{-20px}

\caption{Secondary eclipse and transit depths for Kepler-1~b (TrES-2~b), Kepler-2~b (HAT-P-7~b), Kepler-5~b, Kepler-6~b, Kepler-7~b, and Kepler-8~b. For each target, the secondary eclipse (upper sub-panel) and transit (lower sub-panel) depths by quarter (black) and by 4-quarter window (blue) are displayed. In addition, the depth determined from all quarters stitched together is shown in red, with the $1\sigma$ uncertainty shaded in light red. In the secondary eclipse sub-panels a dotted line is plotted at zero depth, for reference, depending on the plotted relative flux range.\vspace{100px}}
\label{fig:analyses_with_resolved_eclipse_part1}
\end{figure*}
\begin{figure*}[htb!]
\gridline{\fig{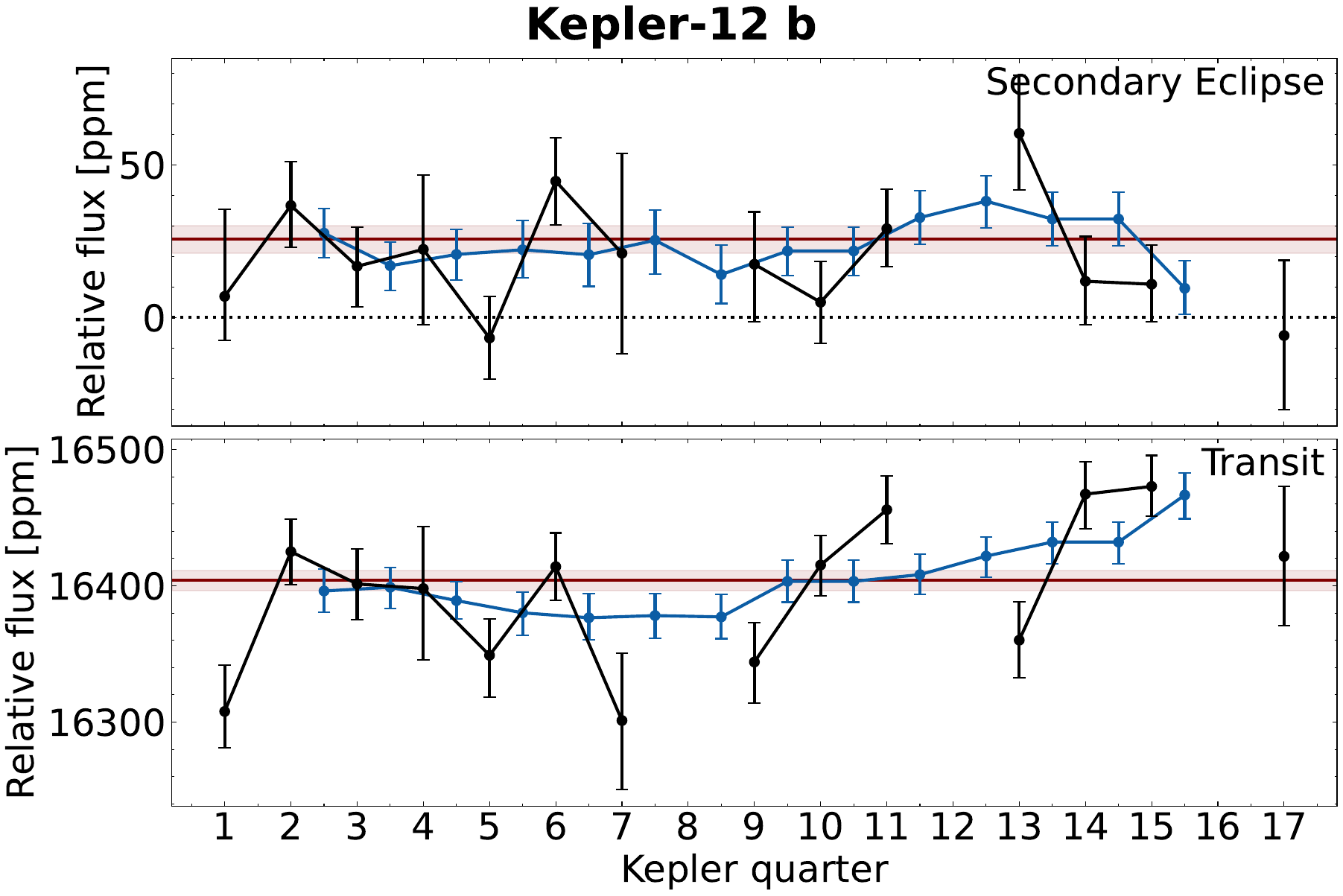}{0.5\textwidth}{}
\fig{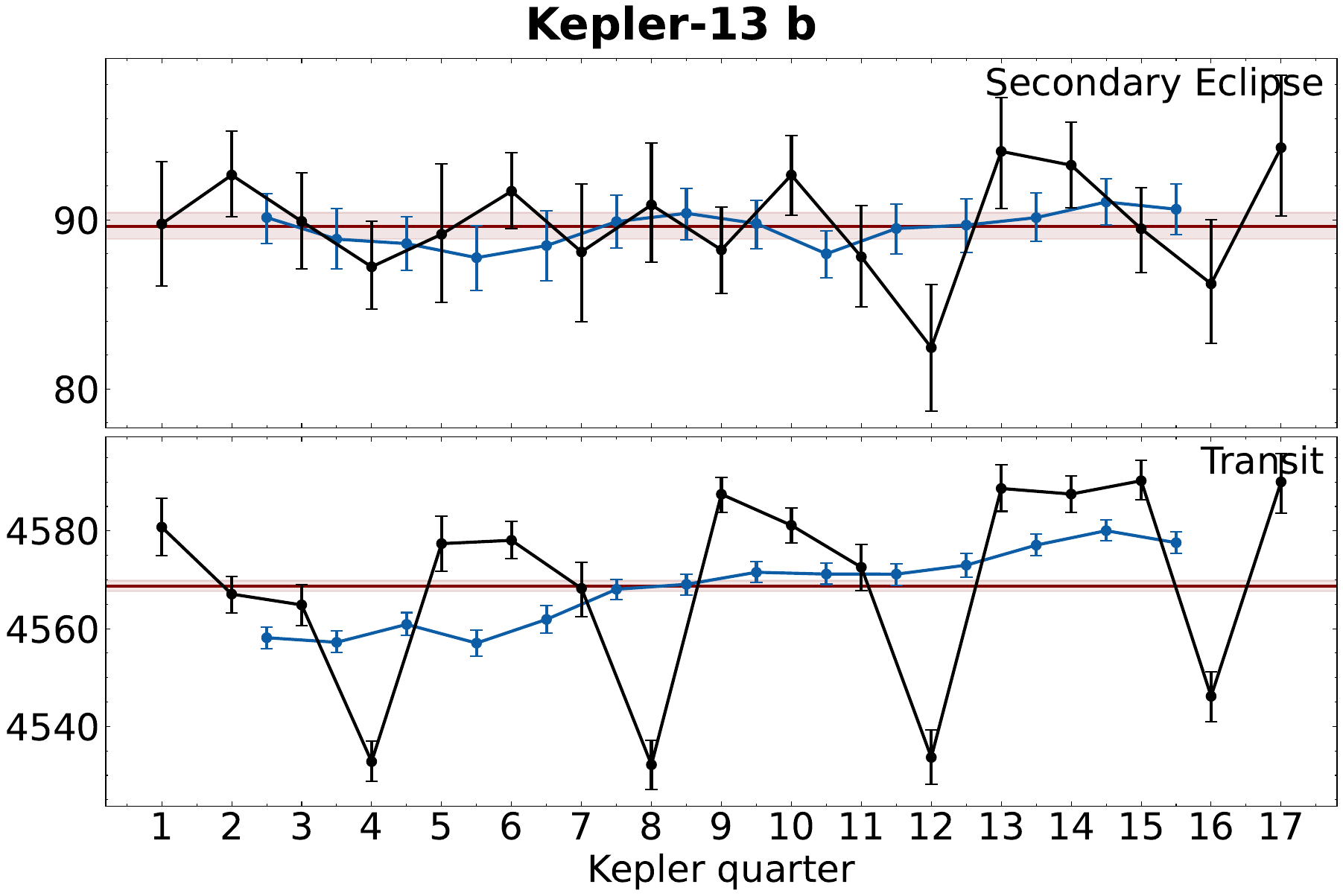}{0.5\textwidth}{}}\vspace{-20px}
\gridline{\fig{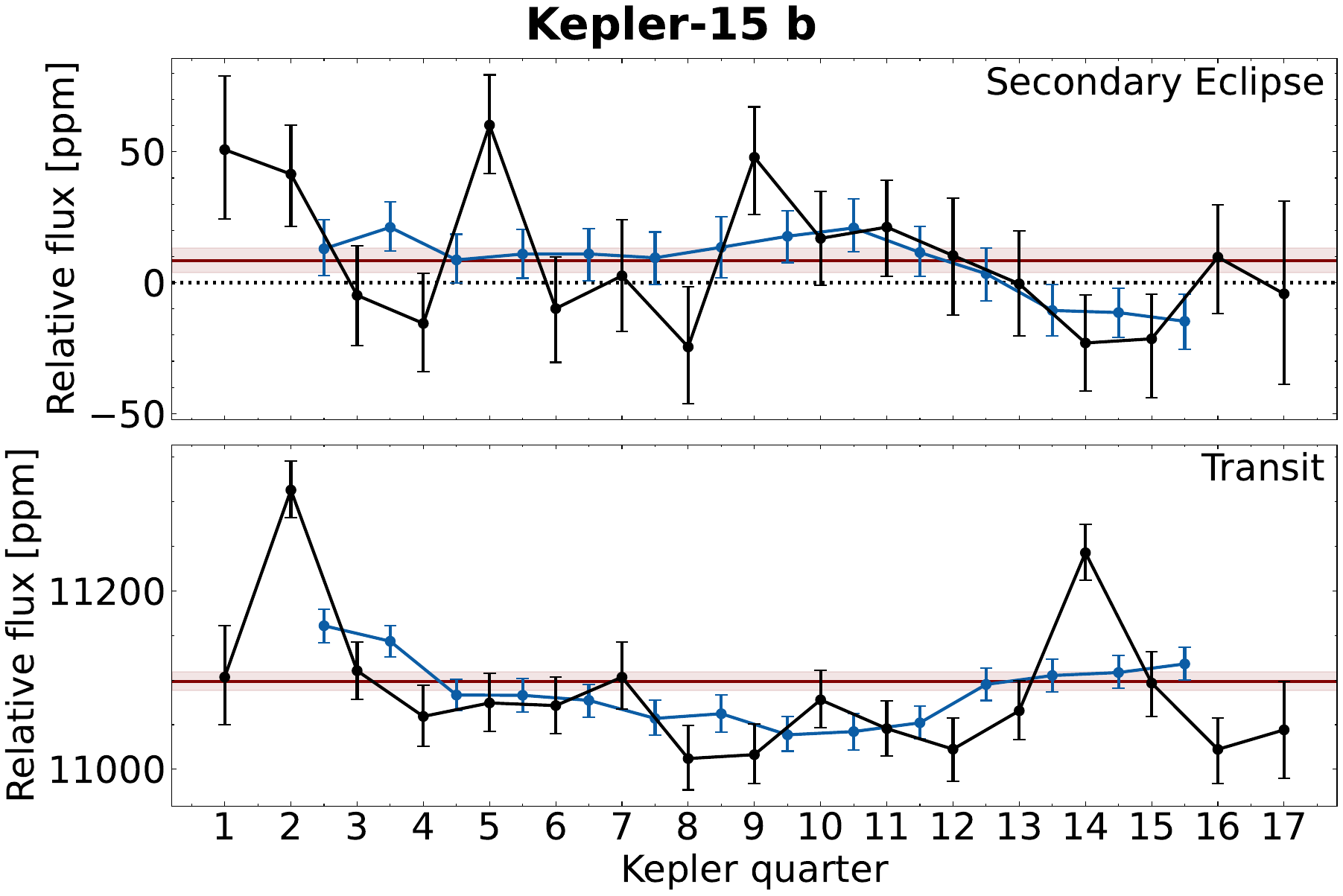}{0.5\textwidth}{}
\fig{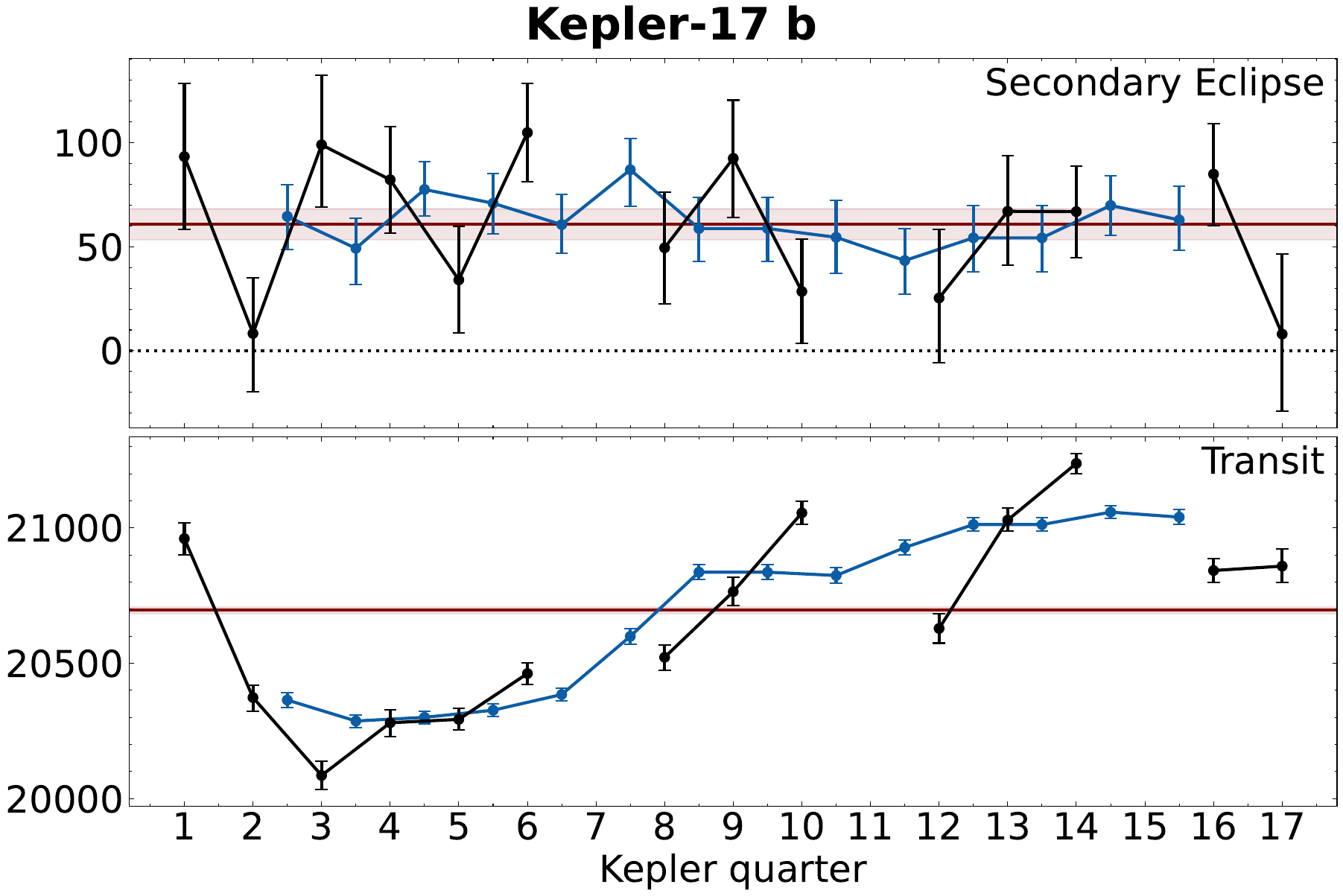}{0.5\textwidth}{}}\vspace{-20px}
\gridline{\fig{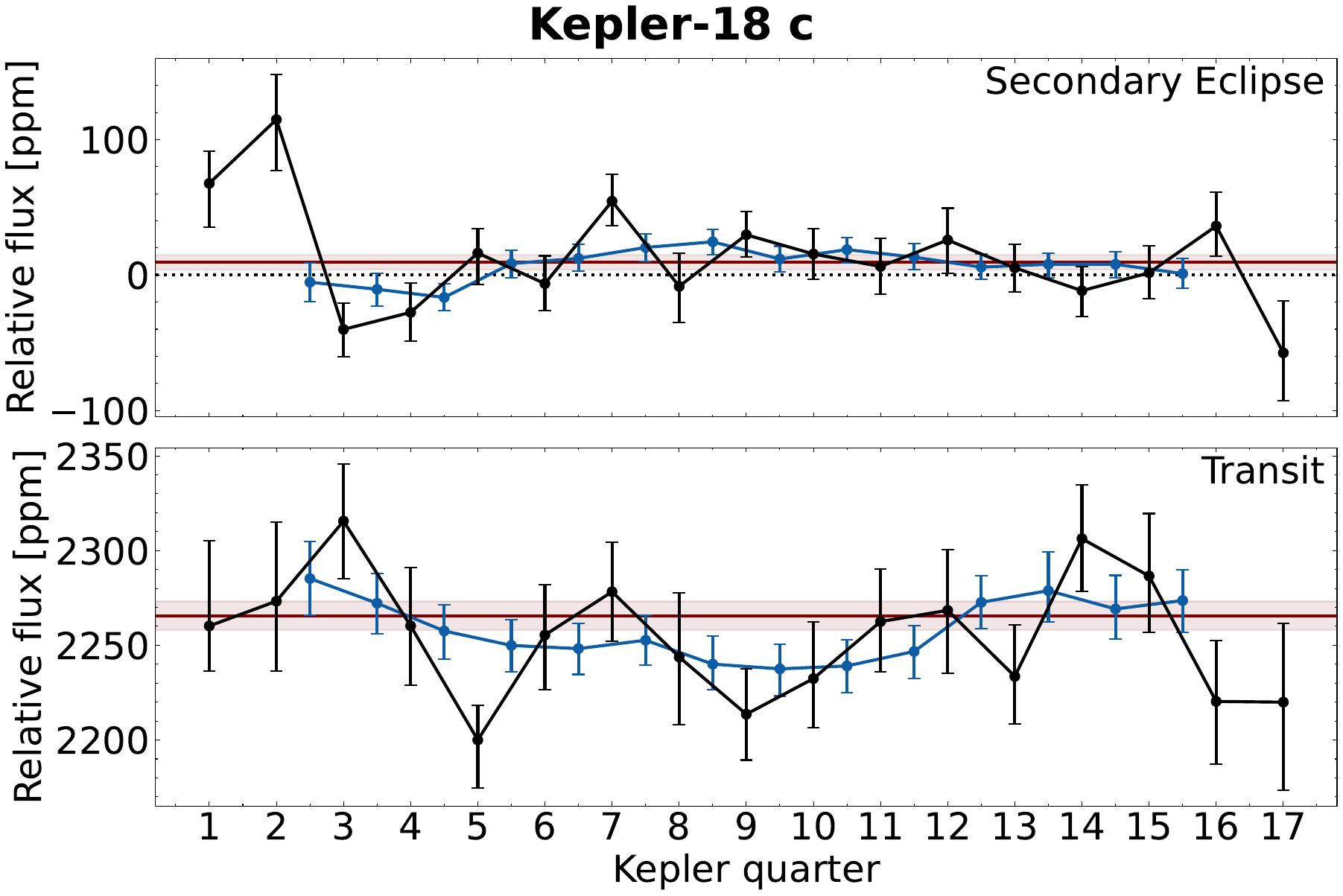}{0.5\textwidth}{}
\fig{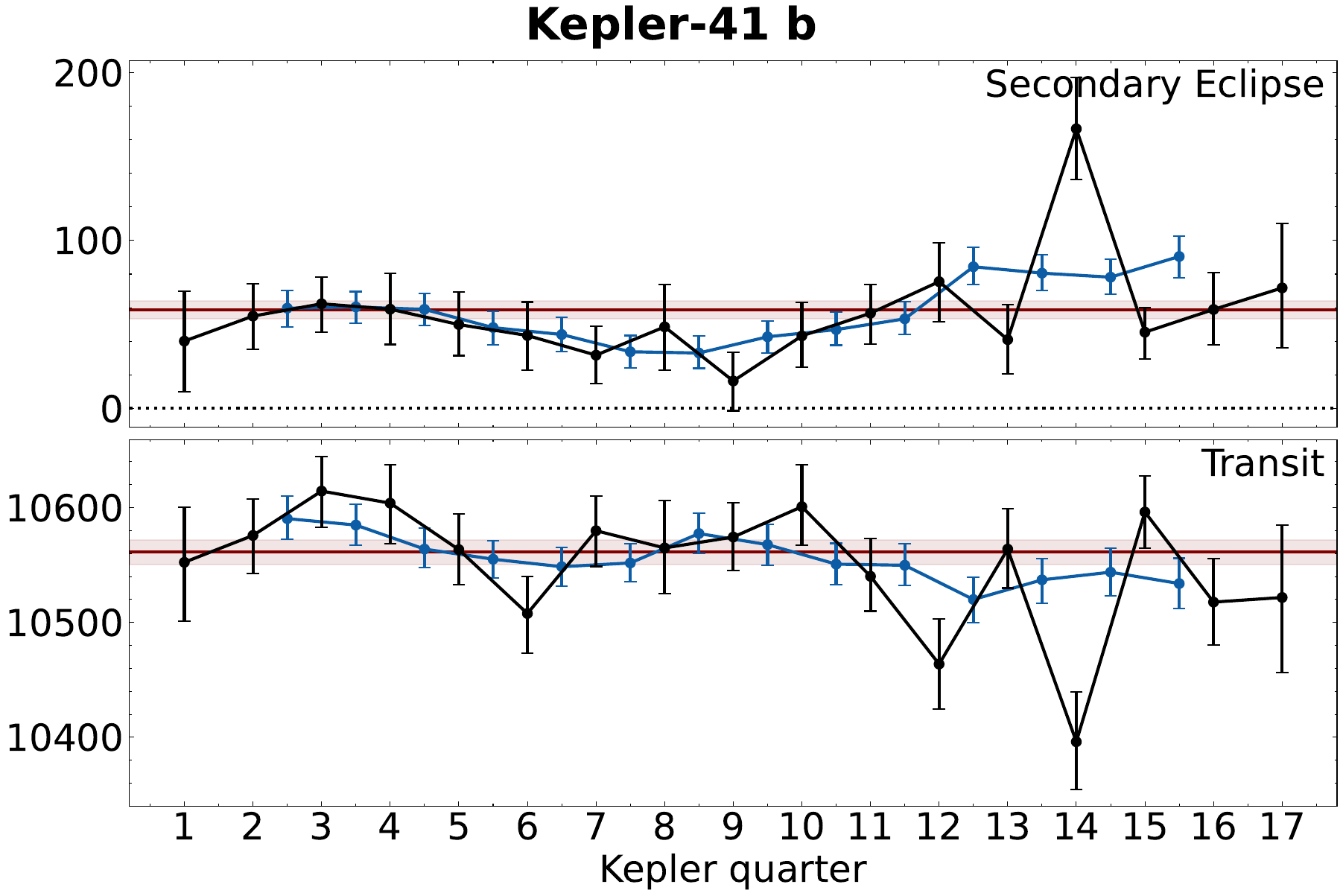}{0.5\textwidth}{}}\vspace{-20px}
\caption{Secondary eclipse and transit depths for Kepler-12~b, Kepler-13~b (KOI-13~b), Kepler-15~b, Kepler-17~b, Kepler-18~c, and Kepler-41~b. Refer to \figr{analyses_with_resolved_eclipse_part1} for details.\vspace{100px}}
\label{fig:analyses_with_resolved_eclipse_part2}
\end{figure*}

\begin{figure*}[htb!]
\gridline{\fig{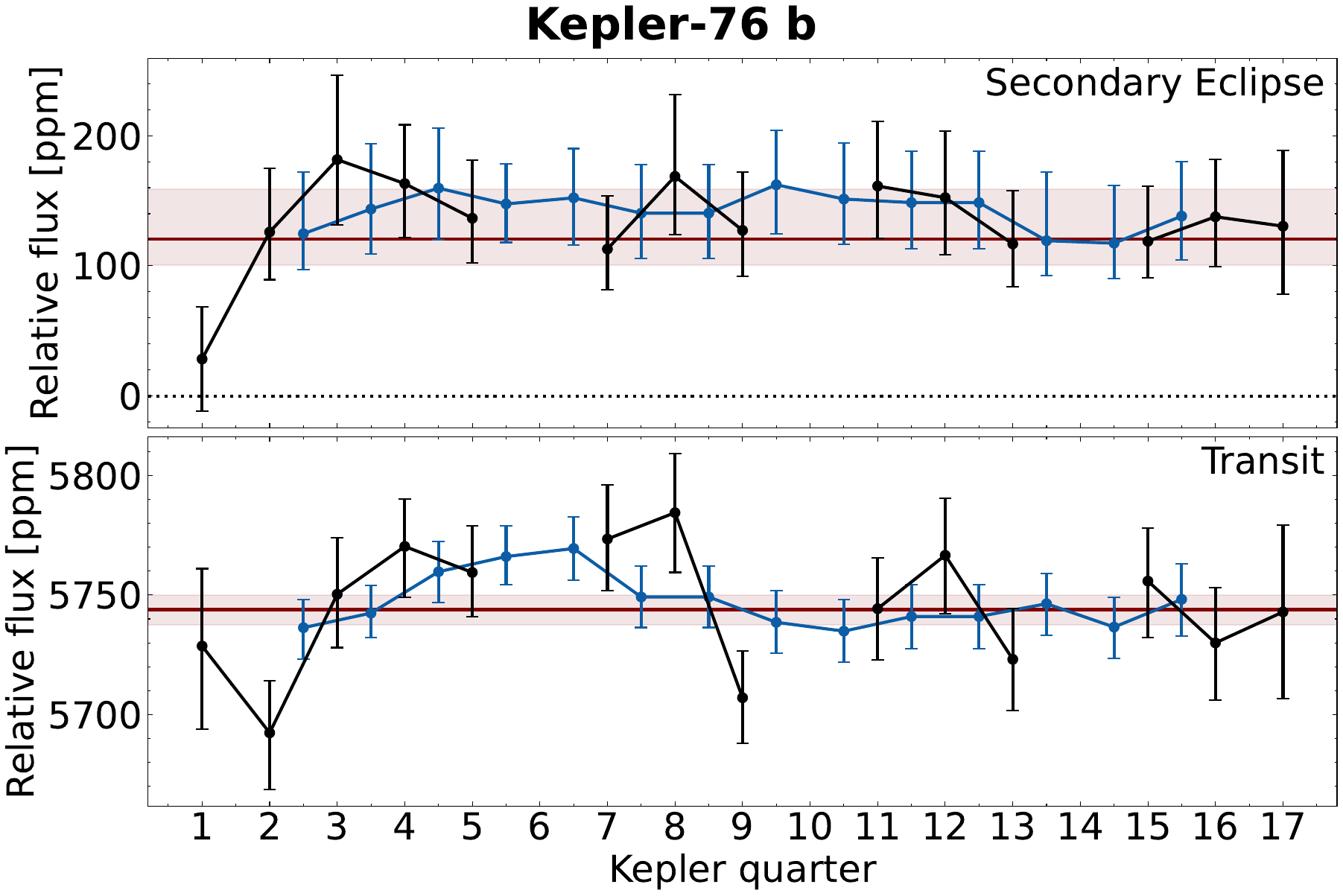}{0.5\textwidth}{}
\fig{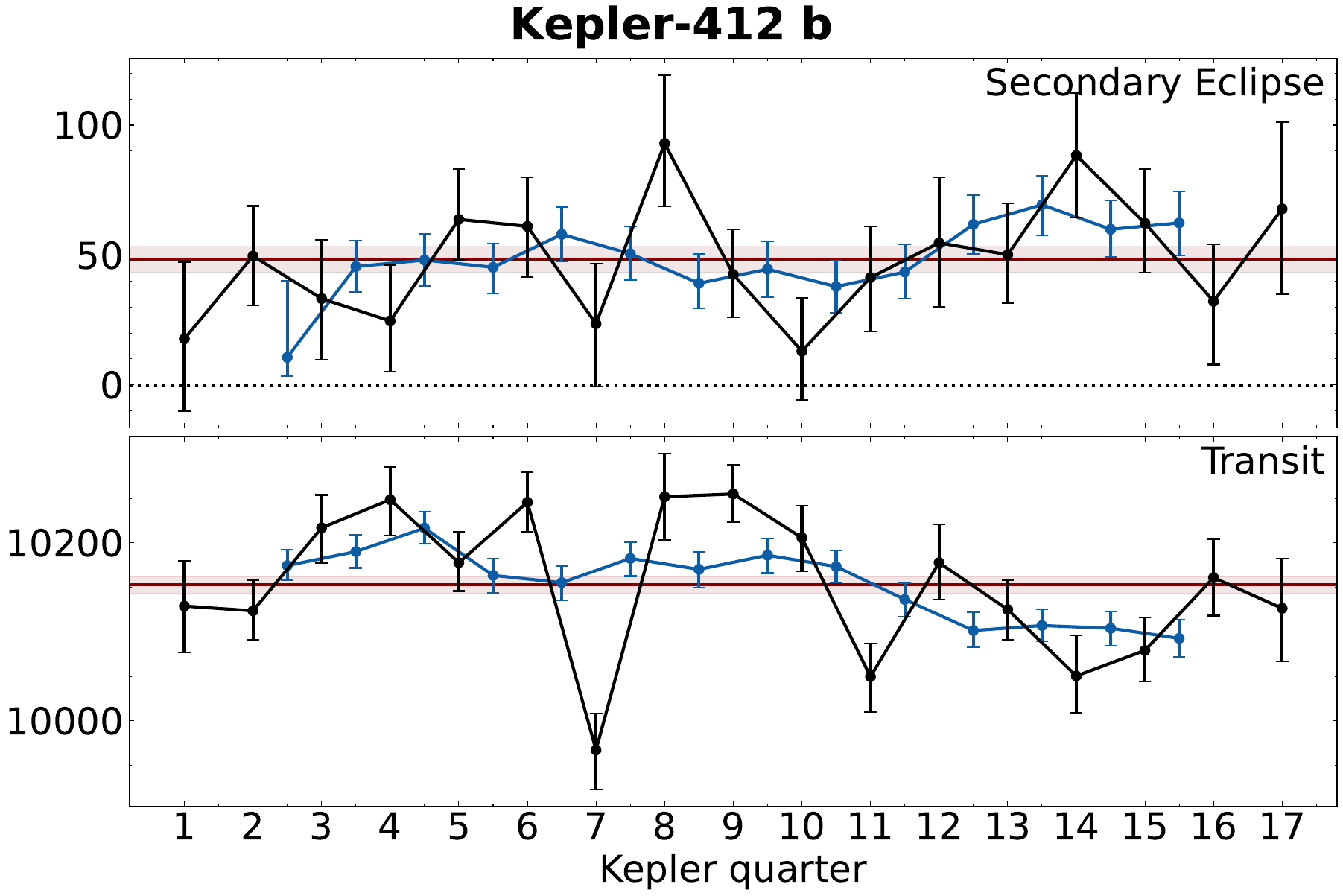}{0.5\textwidth}{}}\vspace{-20px}
\gridline{\fig{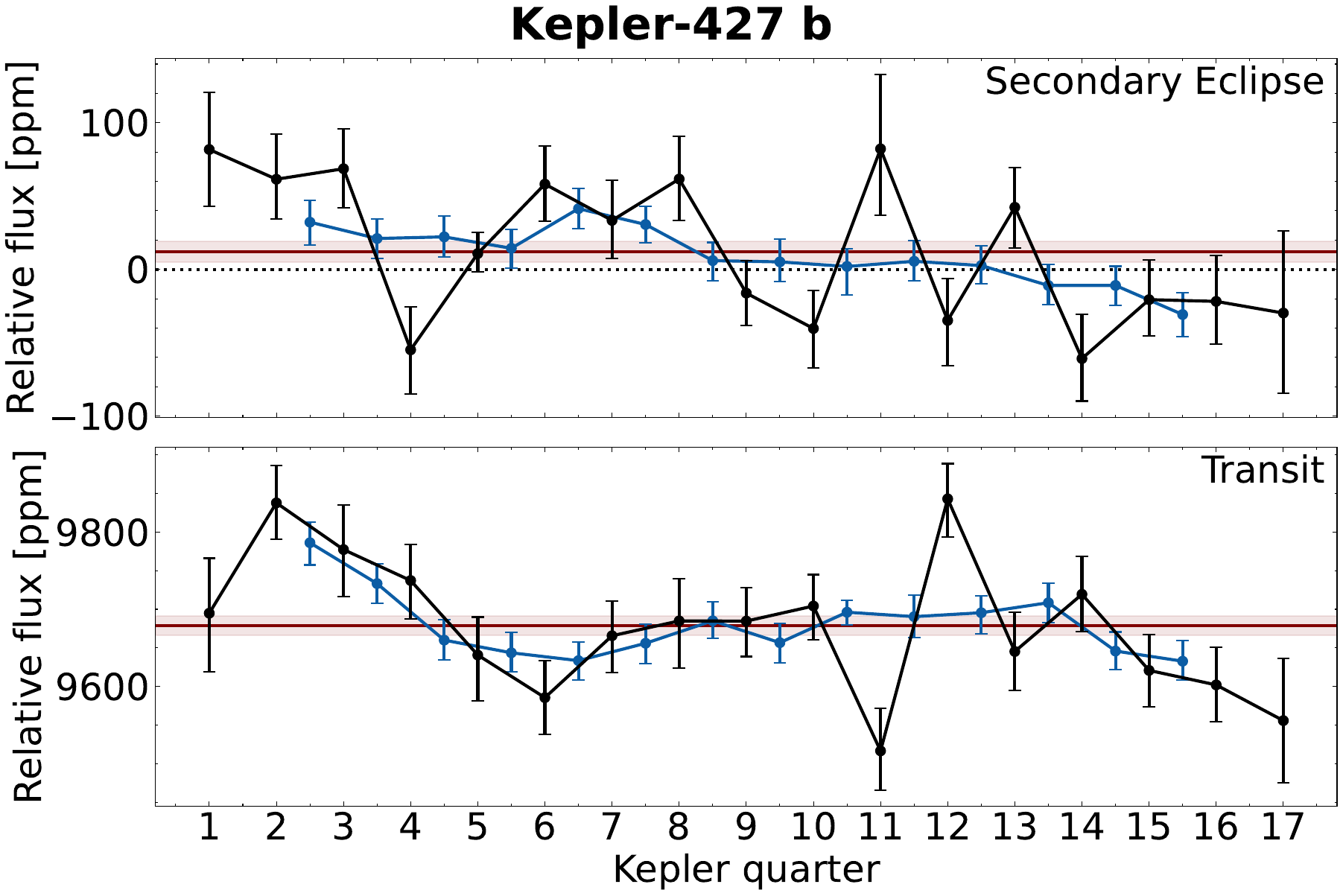}{0.5\textwidth}{}
\fig{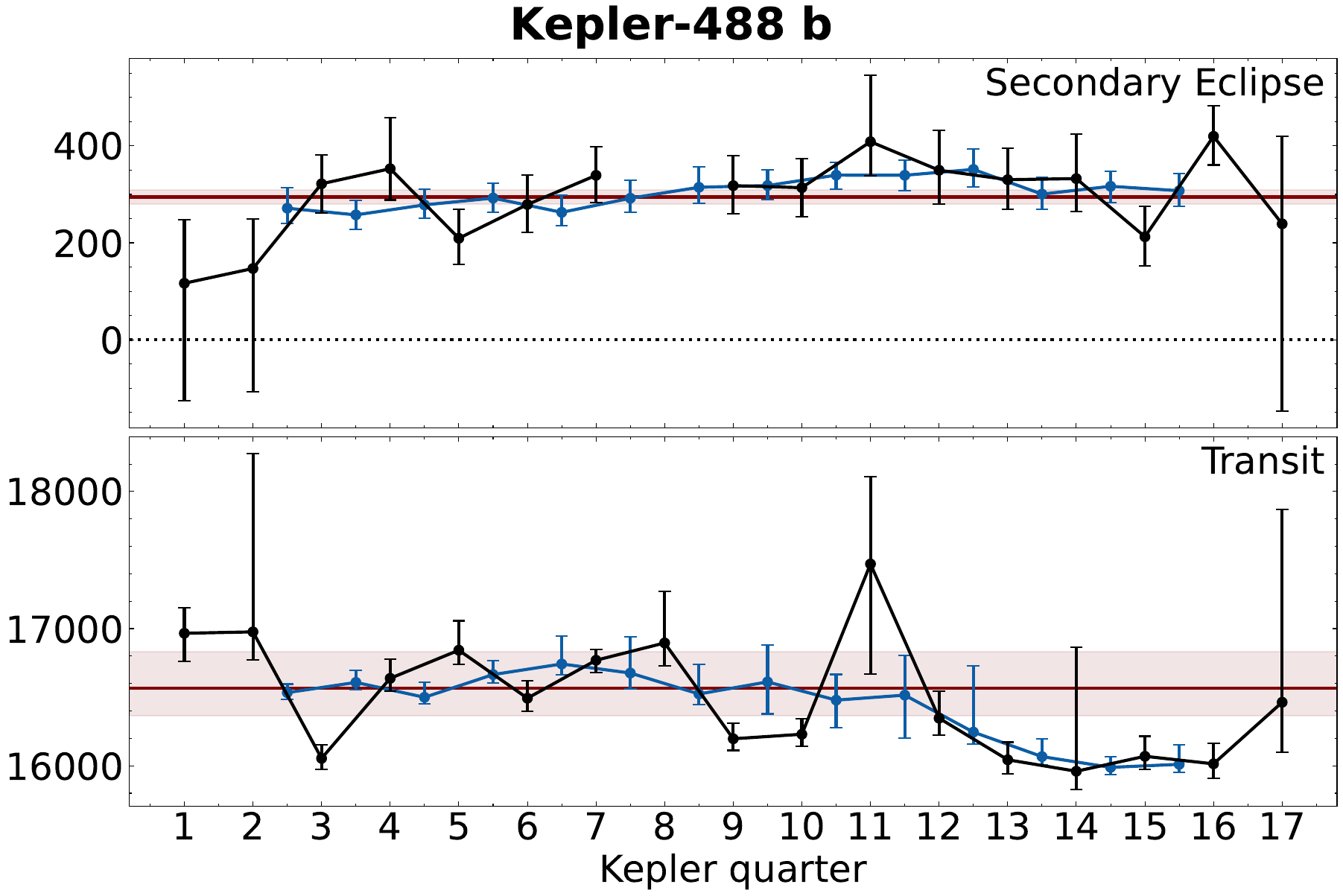}{0.5\textwidth}{}}\vspace{-20px}
\gridline{\fig{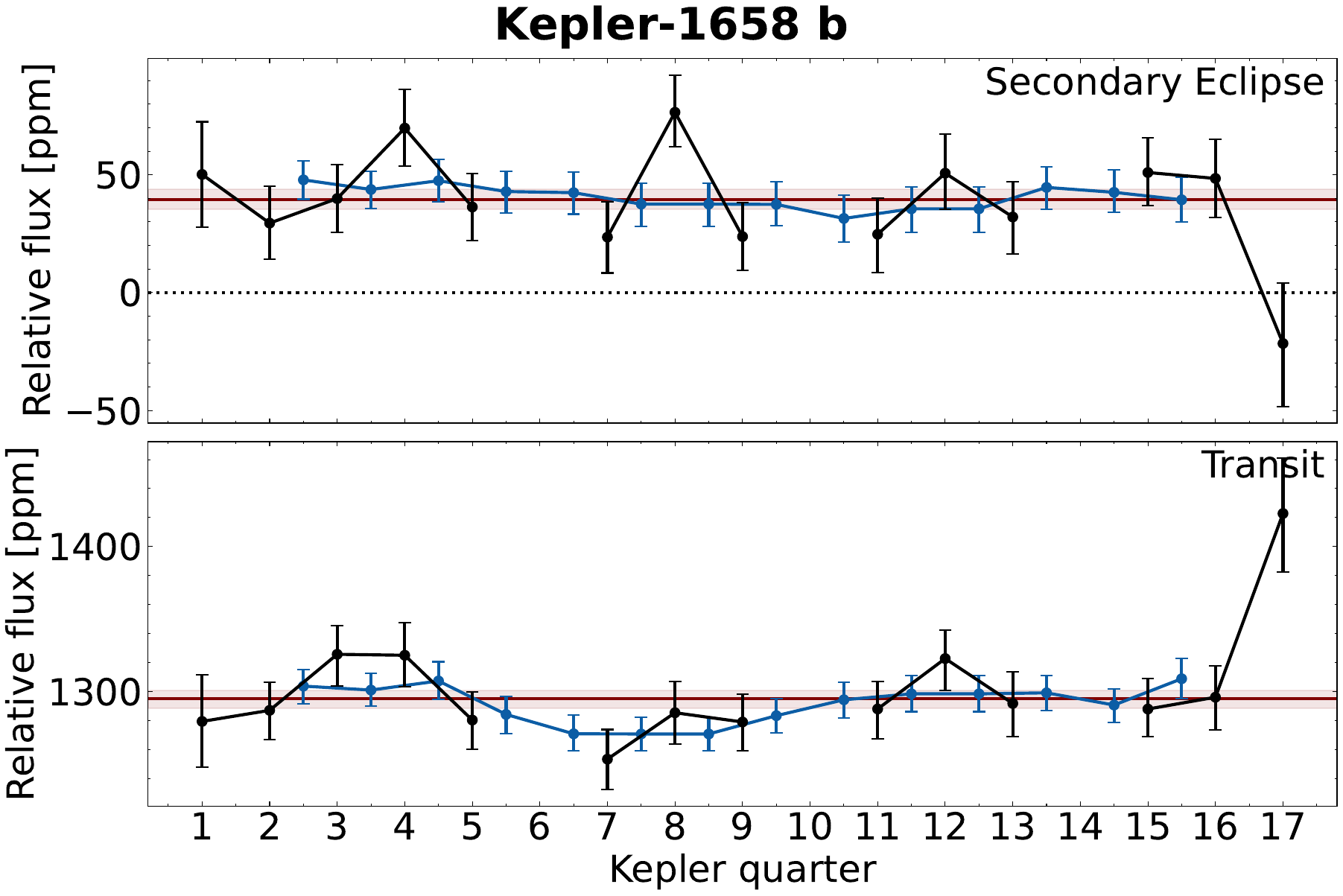}{0.5\textwidth}{}
}\vspace{-20px}
\caption{Secondary eclipse and transit depths for Kepler-76~b, Kepler-412~b, Kepler-427~b, Kepler-488~b, and Kepler-1658~b. Refer to \figr{analyses_with_resolved_eclipse_part1} for details.}
\label{fig:analyses_with_resolved_eclipse_part3}
\end{figure*}

%()()()()()()()()()()()()()()()()()()()()()()()()()()()()()()()()()()()()()()()()

\textbf{Kepler-1~b} (TrES-2~b; \figr{analyses_with_resolved_eclipse_part1} top-left panel):
The quarterly transit depths display a weak 4-quarter (1 year) periodicity. This is similar to the 1-year variability identified by \cite{eylen13} for Kepler-2~b/HAT-P-7~b (see next paragraph) which they named ``seasonal variations''. We adopt that terminology here and discuss it in \secr{seasonvar}.
In addition, the transit depths of the 4-quarter window display a $8.0\sigma$ positive slope of $3.69\substack{+0.44 \\-0.46}$ ppm per quarter, which could be due to orbital precession analogous to what was found for Kepler-13~b by \citet{szabo14, szabo20}. The system was not observed during Q8, Q12, and Q16.

\textbf{Kepler-2~b} (HAT-P-7~b; \figr{analyses_with_resolved_eclipse_part1} top-right panel):
The quarterly transit depths show clear seasonal variations, which seem to also be present in the first half of the quarterly secondary eclipse depth time series but at a low significance and larger relative amplitude. As noted above, this periodicity was already identified by \cite{eylen13}, who concluded that its origin is not astrophysical but instrumental. The correlation between the variations in the quarterly secondary eclipse depths and quarterly transit depths, especially from Q1--Q10, suggests the variability of the former is likely to also originate from the same instrumental process. See \secr{seasonvar} for more details. 

\textbf{Kepler-5~b} (\figr{analyses_with_resolved_eclipse_part1} middle-left panel):
The quarterly secondary eclipse depths show marginal variability in Q2 and Q3, when the depth drops to zero while it is detected at $3.0\sigma$ significance in Q1 and $2.6\sigma$ significance in Q4, and while the transit depths do not show a variation.

\textbf{Kepler-6~b} (\figr{analyses_with_resolved_eclipse_part1} middle-right panel):
The quarterly transit depths display what seems to be seasonal variations, where the depth is at a local maximum in Q6, Q10, and Q14.
Note that the star was not observed during Q7, Q11, and Q15.

\textbf{Kepler-13~b} (KOI-13~b; \figr{analyses_with_resolved_eclipse_part2} top-right panel):
For the transit depths two phenomena are displayed. The first is the instrumental seasonal variation that has a period of one year (4 quarters; see \secr{seasonvar}). The second is a long-term increase in transit depth, seen in the 4-quarter window depths which average out the seasonal variation. This is due to the system's known orbital precession \citep{szabo14, szabo20}. In our analysis of this system we do not account for the dilution due to the system's visual binary companion \citep{shporer14}.

\textbf{Kepler-17~b} (\figr{analyses_with_resolved_eclipse_part2} middle-right panel):
While the data do not show any variability in the secondary eclipse depth, the transit depth displays a large (high signal-to-noise) long-term variation. This star is known to be active, as shown by \citet{desert11}, so the variation in transit depths might be due to varying levels of stellar activity across the 4 years span of the \ik\ data, leading to variations in the non-spotted area of the stellar surface.  
Note that the system was not observed during Q7, Q11, and Q15.

\textbf{Kepler-18~c} (\figr{analyses_with_resolved_eclipse_part2} bottom-left panel):
The secondary eclipse shows marginal variability in Q2 when it deviates from the mean by $2.9\sigma$, while the transit depth in that quarter is similar to that of the rest of the quarters.

\textbf{Kepler-41~b} (\figr{analyses_with_resolved_eclipse_part2} bottom-right panel):
The secondary eclipse depths show an abrupt increase in depth in a single quarter, Q14, compared to the two adjacent quarters. However, the same quarter shows an abrupt drop in transit depth.

\textbf{Kepler-427~b} (\figr{analyses_with_resolved_eclipse_part3} middle-left panel):
While the secondary eclipse depths show a significant downward slope, which we measure to be $-2.44\substack{+0.63 \\-0.74}$ ppm per quarter, we do not consider it as astrophysical since for most quarters the depth is not detected, including low significance measurements and negative depths.

%%%%%%%%%%%%%%%%%%%%%%%%%%%%%%%%%%%%%%%%%%%%%%%%%%%%%%%%%%%%%%%%%%%%%%%%%%%%%
\section{Discussion}
\label{sec:discussion}
%%%%%%%%%%%%%%%%%%%%%%%%%%%%%%%%%%%%%%%%%%%%%%%%%%%%%%%%%%%%%%%%%%%%%%%%%%%%%

We do not observe a statistically significant variability in the quarterly secondary eclipse depths for any of our 53 targets. Taken at face value, the reduced $\chi^2$ values listed in \tabr{kepler_targets_with_resolved_eclipses} for several systems, like Kepler-2~b (HAT-P-7~b), Kepler-18~c, and Kepler-427~b, suggest significant variability, but they are either statistically marginal or instrumental as explained in the previous section.

%For three systems, including Kepler-5~b, Kepler-18~c, and Kepler-41~b, we detect marginal variability during 1--2 quarters. To investigate this further we divided each \ik\ quarter into three equal parts, each with a duration of 30 days, or one month, and then fitted the secondary eclipse and transit to each monthly light curve, giving 51 total measurements of secondary eclipse and transit depths. The results are shown in green in \figr{twelfths}, showing that the monthly measurements are consistent with the depth during the corresponding quarter.

%For those objects additional future observations of the secondary eclipse can be useful. %The appearance of planets where the secondary eclipse is only detected by $3\sigma$ in one quarter (Kepler-15~b, Kepler-18~c) provides confidence that our $3\sigma$ filtering criteria was exhaustive in selecting potential candidates.
%by quarter aside from an outlier in one or two quarters. It is not unusual for the depths during Q1 and Q17 to have a larger error bar from the rest of the quarters due to their shorter duration. 
%For Kepler-8~b and Kepler-41~b, the secondary eclipse depths by 4-quarter windows show a potential long-term variability.

To search for variability on time scales longer than a \ik\ quarter we measured the secondary eclipse depths in 4-quarter windows, and looked for a long-term trend in those depths measurements. None of the systems show a statistically significant trend in the secondary eclipse depths, with the exception of Kepler-427~b. However, as seen in \figr{analyses_with_resolved_eclipse_part3}, the majority of Kepler-427~b's quarterly and 4-quarter secondary eclipse depths are not statistically significant, so the negative slope is likely not astrophysical. 

The lack of systems with clear variability naturally leads to the question: If a system exhibited a change in secondary eclipse depth analogous to that detected for KELT-1~b with TESS data \citep{parviainen23}, would we have been able to detect it in our analysis? \tabr{kepler_targets_with_resolved_eclipses} includes the RMS of the quarterly secondary eclipse depths. All values are well below 100 ppm and most are well below 50 ppm. Those standard deviation estimates in the measured quarterly secondary eclipse depths are much smaller than the $286 \pm 65 $ ppm decrease in depth from TESS Sector 17 to Sector 57 detected by \cite{parviainen23}.

The TESS and \ik\ bandpass are different, with the TESS bandpass extending to redder wavelengths. However, the difference does not significantly affect planetary secondary eclipse depths dominated by light from the host star reflected by the planet's atmosphere, as opposed to thermal emission following heating by the host star. And it is variation in the amount of reflected light that is likely the cause of the variable KELT-1~b secondary eclipse depth reported by \cite{parviainen23}.

Therefore, the non-detection of clear variability in the systems we studied suggests that the variability measured for KELT-1~b is not common, or perhaps unique. This might be related to the fact that KELT-1~b is a brown dwarf with a mass of $27.38\pm0.93\ M_{\rm Jup}$ \citep{siverd12}, and with a surface gravity that is over an order of magnitude larger than that of a typical hot Jupiter \citep{beatty17}. 

%----------------------------------------------------------------------------
\subsection{Seasonal variations in the transit depths}
\label{sec:seasonvar}
%----------------------------------------------------------------------------

A few of the 17 systems presented in \figr{analyses_with_resolved_eclipse_part1}, \figr{analyses_with_resolved_eclipse_part2}, and  \figr{analyses_with_resolved_eclipse_part3}, display a variability in the transit depth with a period of 4 quarters, or 1 year. \citet{eylen13} reported this 4-quarter modulation in Kepler-2~b's (HAT-P-7~b) transit depths by examining the depths of individual transit events up to Q13, and coined them as ``seasonal variations". Our results for Kepler-2~b (\figr{analyses_with_resolved_eclipse_part1} top-right panel) show they persist for the entirety of the \textit{Kepler} data. 

\begin{figure}[tb!]
\gridline{\fig{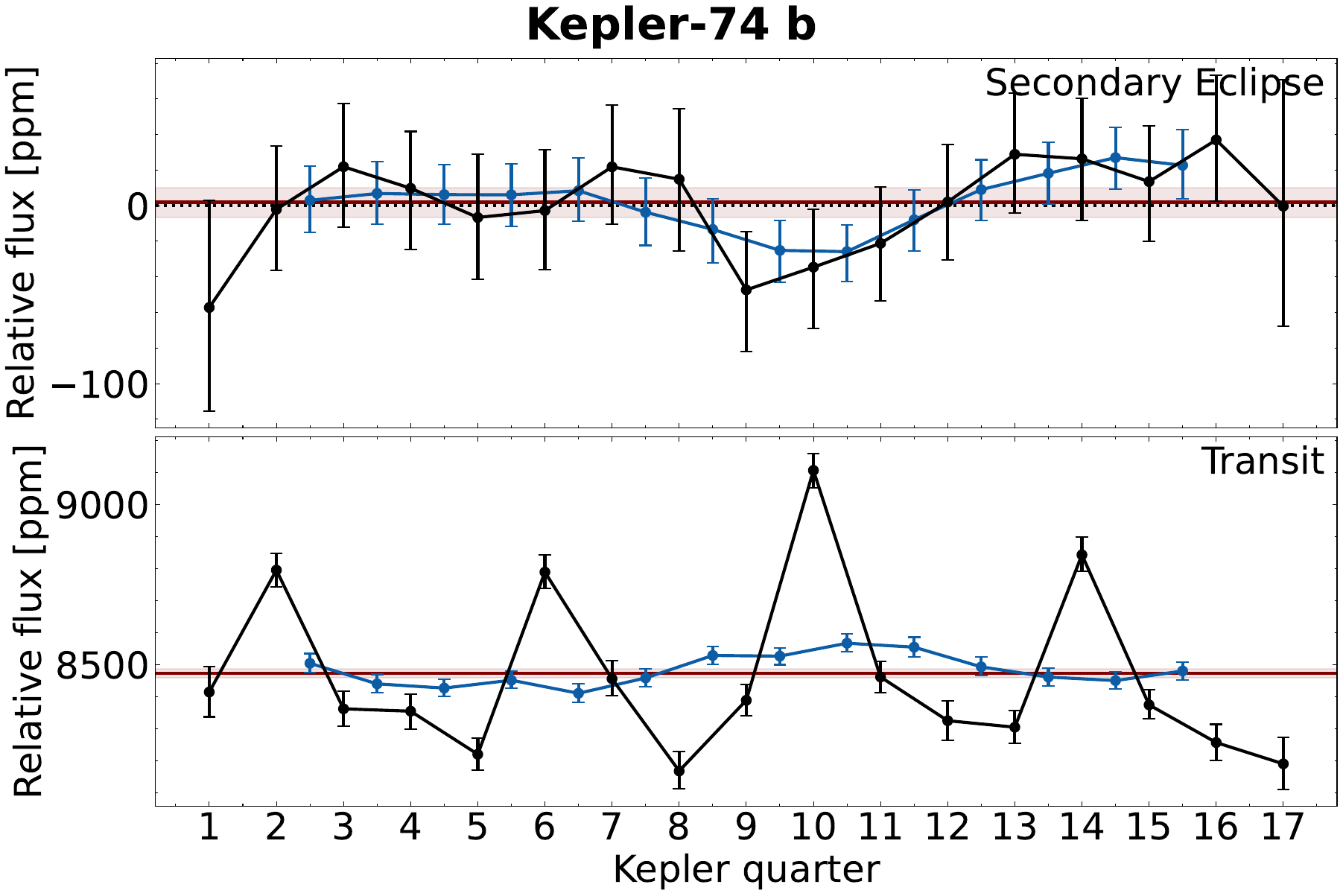}{0.45\textwidth}{}}\vspace{-20px}
\gridline{\fig{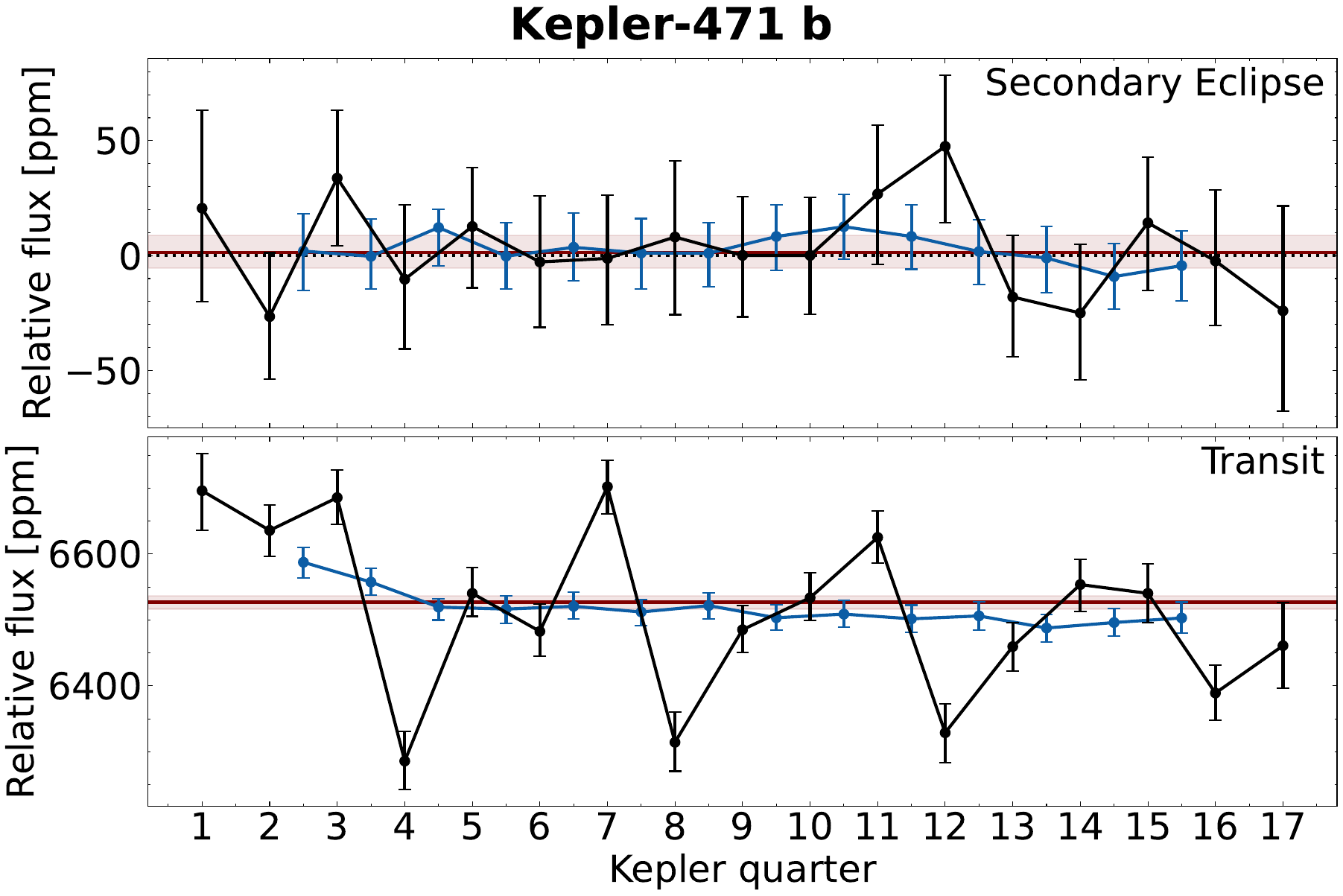}{0.45\textwidth}{}}\vspace{-20px}
\caption{Targets with clear seasonal variations (variability in the transit depth at 1 year, or 4 quarters) not already plotted in \figr{analyses_with_resolved_eclipse_part1} or \figr{analyses_with_resolved_eclipse_part2}. The upper and lower panel show Kepler-74~b and Kepler-471~b, respectively.}
\label{fig:remaining_targets_with_seasonal_variations}
\end{figure}

Moreover, we find similar seasonal variations in the transit depths of Kepler-1~b (TrES-2~b; \figr{analyses_with_resolved_eclipse_part1} top-left panel), Kepler-6~b (\figr{analyses_with_resolved_eclipse_part1} mid-right panel), Kepler-13~b (\figr{analyses_with_resolved_eclipse_part2} bottom-right panel), Kepler-74~b (\figr{remaining_targets_with_seasonal_variations} top panel), and Kepler-471~b (\figr{remaining_targets_with_seasonal_variations} bottom panel). We also looked for this signal in other systems we have analyzed but did not detect the secondary eclipse. Among those, seasonal variations in the transit depth are seen in Kepler-74~b and Kepler-471~b, as shown in \figr{remaining_targets_with_seasonal_variations}.

These seasonal variations are likely caused by an inconsistent blending factor due to unaccounted field crowding or another instrumental effect \citep{eylen13}. Being instrumental in nature, we expect that the secondary eclipses are affected by these seasonal variations at the same amplitude. However, since the observed amplitude of this modulation in the transit depths is on the order of $1\%$, it is not expected to yield a statistically significant variation in the secondary eclipse depths, where the error bars are typically an order of magnitude or larger.

The fact that we see a 4-quarter modulation of the transit depth in several of the systems we analyzed suggests it is present in the light curve of other targets as well, perhaps even many of them where the transit depth is measured at or better than a precision of 1\% in single quarters, as already suggested by \citet{eylen13}. If true, the uncertainties in fitted planet parameters, including that of the planet radius, would be underestimated.

%----------------------------------------------------------------------------
\subsection{Kepler-488~b is a false positive}
\label{sec:kepler488b}
%----------------------------------------------------------------------------

\begin{figure*}[htb!]
\gridline{\fig{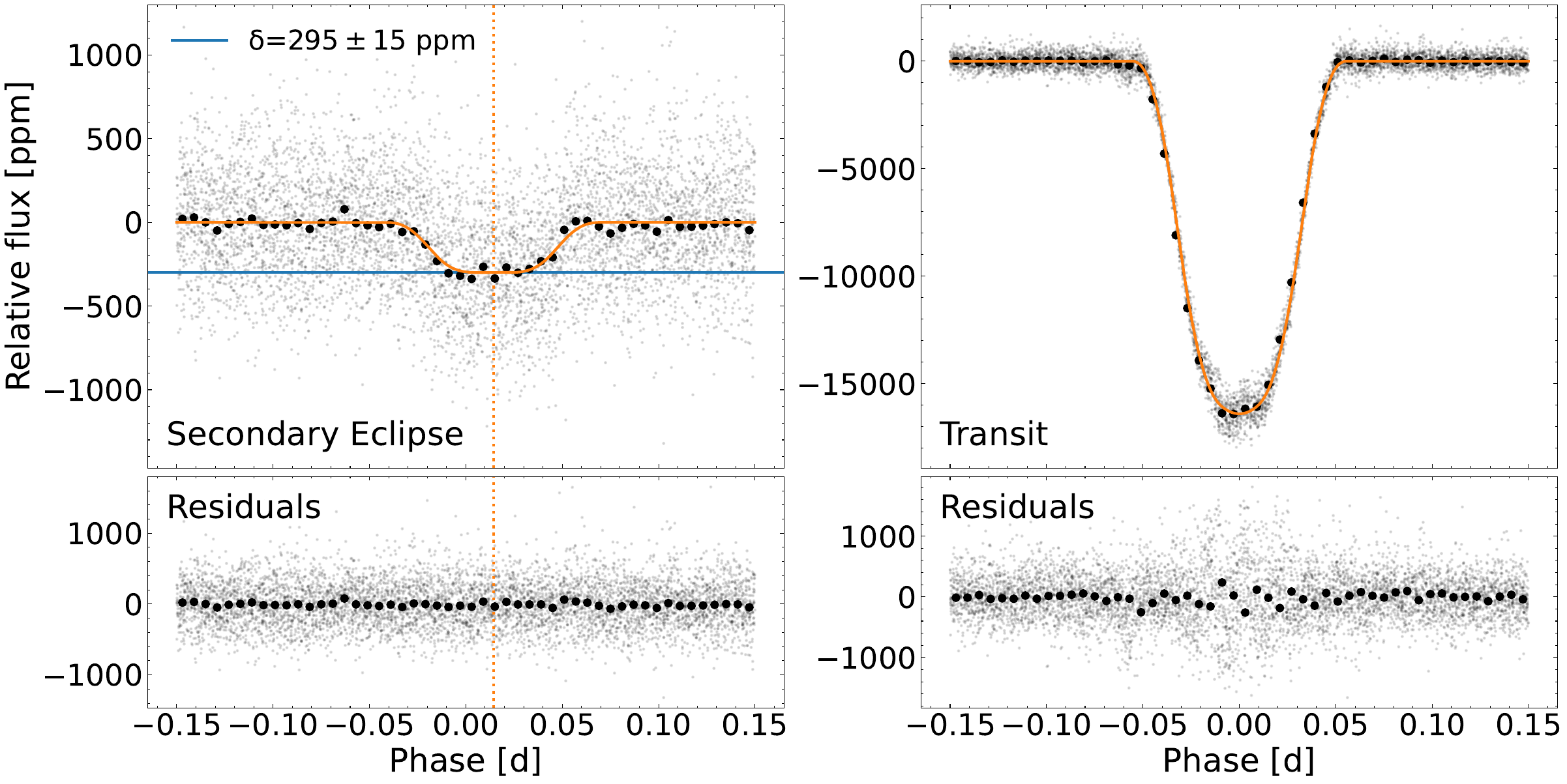}{0.8\textwidth}{}}
\vspace{-20pt}
\caption{Phase-folded light curve of Kepler-488b's secondary eclipse (left) and transit (right), with the model light curve plotted in orange. The long-cadence observations are marked by the gray points in the background, and the solid black circles denote the binned light curve. The nuisance signal modeled by the GP has already been subtracted. The blue line denotes the secondary eclipse depth. The vertical orange dotted line is the center phase of the secondary eclipse, shown to be offset from the expected phase (0.00) assuming a circular orbit (see \secr{kepler488b} for more details). The increase in the scatter of the in-transit residuals (bottom-right panel) is caused by spot-crossing events.}
\label{fig:kepler-488b}
\end{figure*}

Our results for Kepler-488~b (\figr{analyses_with_resolved_eclipse_part2} bottom-right panel) show an unusually deep secondary eclipse, with a depth of about 300 ppm throughout most quarters, deeper than our predicted maximum listed in \tabr{kepler_targets}.
The phase-folded light curve using all available quarters along with our fitted model is shown in \figr{kepler-488b}. 

Note that the phase-folded light curve shows an offset of the secondary eclipse away from half the orbital phase between transits of about 0.015 days. Since that is an order of magnitude larger than the light travel time delay (which is  $2a/c\approx50\;\text{s}$, where $a$ is the orbital semi-major axis and $c$ the speed of light), we assume it is due to a non-zero orbital eccentricity. Fitting for eccentricity results in 
$\sqrt{e}\sin\omega=-0.369\substack{+0.394 \\ -0.080}$ and
$\sqrt{e}\cos\omega=0.0192\substack{+0.0547 \\ -0.0039}$. 
The high statistical significance of the deviation of $\sqrt{e}\cos\omega$ from zero confirms the visual offset, while the consistency of $\sqrt{e}\sin\omega$ with zero means there is no significant difference between the secondary eclipse duration and that of the transit.

Our model fitting of all \ik\ data for Kepler-488~b (\figr{kepler-488b}) resulted in a secondary eclipse depth of $295\pm 15\;\text{ppm}$. 
Assuming that the missing light is entirely composed of reflected light,
the derived $A_g$ in the \ik\ passband using \eqr{albedo} is $1.53\substack{+0.29 \\ -0.30}$. This value is unphysical for hot Jupiters, and even the 3$\sigma$ lower limit is well above measured values of geometric albedos of most hot Jupiters as noted in \secr{targetselection}. Alternatively, assuming the secondary eclipse to be caused solely by thermal emission and taking the temperature of the host star to be $T_*=6134\pm215\;\text{K}$ \citep{berger2018}, the dayside temperature of the planet needed to produce the observed depth would be around $\SI{2900}{K}$, accounting for the \ik\ passband. This surface temperature corresponds to that of an M5/M6 red dwarf. Such dayside temperature is significantly greater than the planet's dayside equilibrium temperature, computed to be $1657\substack{+114 \\ -89}\;\text{K}$ using \eqr{T_eq_dayside}. Since neither missing light component can account for its large secondary eclipse depth, Kepler-488~b cannot be a planet and must be self-luminous. 

To look for similar cases of deep secondary eclipses leading to unphysical albedos, we use \eqr{albedo} to estimate the geometric albedo (neglecting thermal emission) of all planets in \tabr{kepler_targets_with_resolved_eclipses}. 
None of the resulting geometric albedo values, besides that of Kepler-488~b, are above 1.0. For the planets where we were able to find a published geometric albedo, the values we calculated are statistically consistent with the literature (\citet{angerhausen15}, \citet{esteves15}, \citet{chontos19}). 
% The other target we find to have a large, statistically significant geometric albedo is Kepler-1658b, with $A_g=0.751\substack{+0.099\\-0.126}$. This value is consistent with the case study by \citet{chontos19}, who confirmed its planetary nature and determined the albedo to be $0.724\substack{+0.090\\-0.081}$.

%%%%%%%%%%%%%%%%%%%%%%%%%%%%%%%%%%%%%%%%%%%%%%%%%%%%%%%%%%%%%%%%%%%%%%%%%%%%%
\section{Summary}
\label{sec:summary}
%%%%%%%%%%%%%%%%%%%%%%%%%%%%%%%%%%%%%%%%%%%%%%%%%%%%%%%%%%%%%%%%%%%%%%%%%%%%%

We present a systematic search for atmospheric variability in short-period gas-giant exoplanets (hot Jupiters) observed by the \ik\ Mission, by looking for variability in the secondary eclipse depths in their \ik\ light curves. Our analysis included fitting the secondary eclipse and transit per quarter for 53 systems that satisfied our selection criteria (\tabr{kepler_targets}). We were able to detect the secondary eclipse with at least $3\sigma$ significance in at least one \ik\ quarter or four-quarter window for 17 out of the 53 systems. 
We did not detect clear, statistically significant non-instrumental variability in the secondary eclipse depth in any of the 17 systems. 

The non-detection, combined with the sensitivity of our analysis to variability at the level of 100 ppm, suggests that exoplanet atmospheric variability with an amplitude observed in KELT-1~b by \citet{parviainen23} is not common. 

While it was not the goal of this work, we detected seasonal variations --- variability with a period of 1 year or four \ik\ quarters --- in the transit depths of Kepler-1~b (TrES-1~b), Kepler-2~b (HAT-P-7~b), Kepler-6~b, Kepler-13~b (KOI-13~b), Kepler-74~b, and Kepler-471~b. The exact origin of this instrumental signal, reported in the past for Kepler-2~b (HAT-P-7~b) by \cite{eylen13}, is not completely clear, and it might be affecting the transit depths of other systems we did not analyze as well as the variability amplitude of other variable stars.

Finally, we showed that Kepler-488~b's secondary eclipse depth is too deep to be explained by planetary thermal emission or reflected light, and it must be a self-luminous object. Therefore, Kepler-488~b is a star, likely a red dwarf, that was misclassified as a planet.

% \begin{acknowledgments}
% Ian Wong and Andrew Vanderburg
We are grateful to Ian Wong and Andrew Vanderburg for providing detailed comments on an early draft of this manuscript. 
% Anonymous Referee
We would also like to thank the anonymous referee for their helpful comments.
% Kepler
This paper includes data collected by the \ik\ mission and obtained from the MAST data archive at the Space Telescope Science Institute (STScI). Funding for the \ik\ mission is provided by the NASA Science Mission Directorate. STScI is operated by the Association of Universities for Research in Astronomy, Inc., under NASA contract NAS 5–26555.
% NASA Exoplanet Archive
This research has made use of the NASA Exoplanet Archive, which is operated by the California Institute of Technology, under contract with the National Aeronautics and Space Administration under the Exoplanet Exploration Program.
% \end{acknowledgments}

\facility{\textit{{Kepler}}, TESS, MAST, Exoplanet Archive}.
\software{exoplanet \citep{exoplanet}, PyMC3 \citep{pymc3}, celerite2 \citep{celerite2}, lightkurve \citep{lightkurve}, numpy \citep{numpy}, matplotlib \citep{matplotlib}, astropy \citep{astropy1, astropy2}}.

\appendix
\setcounter{table}{0}
\renewcommand{\thetable}{A\arabic{table}}

\centering
\begin{longtable}{lrrrr}
\caption{Table of 53 \textit{Kepler} planets selected for our secondary eclipse variability search, including planets with period $< 15$~d, radius $>5R_\oplus$, and $Kp < 15.0$ mag. The rightmost column lists the estimated maximum secondary eclipse depth due to reflected light, assuming a visible geometric albedo of 1.0 (refer to \secr{targetselection}). The content of this table is based on data provided by the NASA Exoplanet Archive.}
\label{tab:kepler_targets}\\

\hline 
\multicolumn{1}{c}{Planet} & \multicolumn{1}{r}{Period} & \multicolumn{1}{r}{Radius} & \multicolumn{1}{c}{$Kp$} & \multicolumn{1}{c}{$D_{\text{max}}$} \\ 
\multicolumn{1}{c}{name}   & \multicolumn{1}{c}{[d]}    & \multicolumn{1}{c}{[$R_\oplus$]} & \multicolumn{1}{c}{[mag]} & \multicolumn{1}{c}{[ppm]} \\ 
\hline 
\endfirsthead

\multicolumn{5}{c}%
{{\bfseries \tablename\ \thetable{} -- Continued from previous page}} \\ 
\hline 
\multicolumn{1}{c}{Planet} & \multicolumn{1}{r}{Period} & \multicolumn{1}{r}{Radius}       & \multicolumn{1}{c}{$Kp$} & \multicolumn{1}{c}{$D_{\text{max}}$} \\ 
\multicolumn{1}{c}{name}   & \multicolumn{1}{c}{[d]}    & \multicolumn{1}{c}{[$R_\oplus$]} & \multicolumn{1}{c}{[mag]} & \multicolumn{1}{c}{[ppm]} \\ 
\hline 
\endhead

\hline \multicolumn{5}{r}{{Continued on next page}} \\ \hline
\endfoot

\hline 
\endlastfoot
% using DR24
Kepler-1~b & 2.47 & 15.2 & 11.34 & $263\substack{+22\\-19}$ \\
Kepler-2~b & 2.20 & 16.9 & 10.46 & $334\substack{+26\\-23}$ \\
Kepler-5~b & 3.55 & 16.0 & 13.37 & $153.7\substack{+1.2\\-1.1}$ \\
Kepler-6~b & 3.23 & 14.6 & 13.30 & $157.8\substack{+1.0\\-1.0}$ \\
Kepler-7~b & 4.89 & 18.2 & 12.88 & $156.2\substack{+1.1\\-1.1}$ \\
Kepler-8~b & 3.52 & 15.9 & 13.56 & $195.1\substack{+1.3\\-1.3}$ \\
Kepler-12~b & 4.44 & 19.7 & 13.44 & $219.7\substack{+0.8\\-0.8}$ \\
Kepler-13~b & 1.76 & 16.9 & 9.96 & $376.8\substack{+0.7\\-0.7}$ \\
Kepler-14~b & 6.79 & 12.7 & 12.13 & $46.8\substack{+3.3\\-5.3}$ \\
Kepler-15~b & 4.94 & 10.8 & 13.76 & $61\substack{+17\\-10}$ \\
Kepler-17~b & 1.49 & 14.7 & 14.14 & $565.5\substack{+4.5\\-4.4}$ \\
Kepler-18~c & 7.64 & 5.5 & 13.55 & $10.0\substack{+0.9\\-0.8}$ \\
Kepler-18~d & 14.86 & 7.0 & 13.55 & $6.6\substack{+0.6\\-0.6}$ \\
Kepler-25~c & 12.72 & 5.2 & 10.73 & $0.90\substack{+0.02\\-0.02}$ \\
Kepler-40~b & 6.87 & 13.1 & 14.59 & $48.5\substack{+1.0\\-1.0}$ \\
Kepler-41~b & 1.86 & 14.5 & 14.46 & $395.9\substack{+4.9\\-4.7}$ \\
Kepler-43~b & 3.02 & 13.7 & 13.96 & $152.9\substack{+2.3\\-2.2}$ \\
Kepler-44~b & 3.25 & 12.2 & 14.68 & $137\substack{+16\\-13}$ \\
Kepler-56~b & 10.50 & 6.5 & 12.44 & $3.5\substack{+3.2\\-1.4}$ \\
Kepler-63~b & 9.43 & 6.1 & 11.58 & $10.6\substack{+0.4\\-0.3}$ \\
Kepler-74~b & 7.34 & 10.8 & 14.41 & $34.8\substack{+1.1\\-1.0}$ \\
Kepler-76~b & 1.54 & 15.2 & 13.31 & $534\substack{+28\\-31}$ \\
Kepler-77~b & 3.58 & 10.8 & 13.94 & $103.3\substack{+1.3\\-1.3}$ \\
Kepler-91~b & 6.25 & 15.3 & 12.50 & $76.5\substack{+4.5\\-4.2}$ \\
Kepler-101~b & 3.49 & 5.8 & 13.77 & $26.8\substack{+7.0\\-6.1}$ \\
Kepler-122~c & 12.47 & 5.9 & 14.25 & $6.1\substack{+6.2\\-2.5}$ \\
Kepler-412~b & 1.72 & 15.0 & 14.31 & $467.7\substack{+7.0\\-7.8}$ \\
Kepler-418~c & 12.22 & 5.2 & 14.70 & $1.7\substack{+1.9\\-0.7}$ \\
Kepler-422~b & 7.89 & 12.9 & 13.44 & $40.2\substack{+2.8\\-2.5}$ \\
Kepler-423~b & 2.68 & 13.4 & 14.29 & $241.8\substack{+16.2\\-7.2}$ \\
Kepler-424~b & 3.31 & 10.0 & 14.26 & $74.5\substack{+10.4\\-7.3}$ \\
Kepler-425~b & 3.80 & 11.0 & 14.74 & $101.4\substack{+2.0\\-2.0}$ \\
Kepler-426~b & 3.22 & 12.2 & 14.84 & $159.4\substack{+3.9\\-4.0}$ \\
Kepler-427~b & 10.29 & 13.8 & 14.22 & $41\substack{+16\\-10}$ \\
Kepler-433~b & 5.33 & 16.2 & 14.46 & $105\substack{+23\\-18}$ \\
Kepler-434~b & 12.87 & 12.7 & 14.52 & $22.8\substack{+8.9\\-5.9}$ \\
Kepler-435~b & 8.60 & 22.3 & 13.64 & $101\substack{+18\\-14}$ \\
Kepler-447~b & 7.79 & 18.5 & 12.39 & $65\substack{+45\\-33}$ \\
Kepler-471~b & 5.01 & 15.0 & 13.80 & $71.7\substack{+0.2\\-0.2}$ \\
Kepler-485~b & 3.24 & 14.4 & 14.95 & $174.8\substack{+0.2\\-0.2}$ \\
Kepler-488~b & 3.12 & 15.8 & 14.80 & $216\substack{+19\\-16}$ \\
Kepler-490~b & 3.27 & 11.6 & 14.88 & $142.9\substack{+2.1\\-2.1}$ \\
Kepler-491~b & 4.23 & 8.9 & 14.01 & $41.2\substack{+0.2\\-0.2}$ \\
Kepler-493~b & 3.00 & 15.1 & 15.00 & $135.9\substack{+5.2\\-4.7}$ \\
Kepler-494~b & 8.03 & 7.1 & 14.15 & $19.3\substack{+4.3\\-3.1}$ \\
Kepler-495~b & 3.41 & 5.2 & 14.62 & $30.4\substack{+0.5\\-0.5}$ \\
Kepler-497~b & 3.57 & 6.0 & 14.72 & $26.1\substack{+1.2\\-1.1}$ \\
Kepler-546~b & 4.15 & 7.0 & 14.29 & $23.9\substack{+0.8\\-0.8}$ \\
Kepler-548~b & 4.45 & 12.0 & 15.00 & $67.9\substack{+9.2\\-7.8}$ \\
Kepler-975~c & 5.06 & 10.7 & 14.44 & $21\substack{+63\\-14}$ \\
Kepler-1004~b & 5.29 & 6.3 & 13.43 & $16.1\substack{+30.6\\-8.1}$ \\
Kepler-1517~b & 5.55 & 9.8 & 12.25 & $12.6\substack{+25.4\\-6.6}$ \\
Kepler-1658~b & 3.85 & 12.0 & 11.43 & $83.5\substack{+8.2\\-7.5}$ \\
\end{longtable}
\end{document}